\documentclass[12pt]{article}
\usepackage{amsmath}
\usepackage{graphicx}
\usepackage{enumerate}
\usepackage{amsfonts}
\usepackage{url} 
\usepackage{bbm}
\usepackage{multirow}
\usepackage{makecell}
\usepackage{float}
\usepackage{mathtools}
\usepackage{multibib}
\newcites{latex}{Reference}%

\usepackage{natbib}

\newcommand{\blind}{1}

\newtheorem{definition}{Definition}
\newtheorem{assumption}{Assumption}

\newtheorem{theorem}{Theorem}
\newtheorem{lemma}{Lemma}
\newtheorem{condition}{Condition}
\newcommand{\ind}{\perp\!\!\!\!\perp} 

\DeclareMathOperator*{\argmin}{arg\,min}
\DeclarePairedDelimiter\abs{\lvert}{\rvert}%

\begin{document}

\def\spacingset#1{\renewcommand{\baselinestretch}%
{#1}\small\normalsize} \spacingset{1}


\if1\blind
{
  \title{\bf Estimating treatment effects from observational data under truncation by death using survival-incorporated quantiles}
  \author{Qingyan Xiang\textsuperscript{1}, Paola Sebastiani\textsuperscript{2}, Thomas T. Perls\textsuperscript{3}, \\ Stacy L. Andersen\textsuperscript{3}, Svetlana Ukraintseva\textsuperscript{4}, \\Mikael Thinggaard\textsuperscript{5}, and Judith J. Lok\textsuperscript{6} \thanks{
    This work was sponsored by NSF grant DMS 1854934 to Judith J. Lok and NIA grant U19AG063893 to all Long Life Family Study investigators. This paper constitutes the second part of Qingyan Xiang's PhD dissertation, and the authors would like to thank his dissertation committee members, Ronald J. Bosch, Sara Lodi, and Yorghos Tripodis, for their invaluable input. The authors would like to thank all Long Life Family Study participants for their involvement.}\hspace{.2cm}\\
    \\
    1 Department of Biostatistics, Vanderbilt University Medical Center \\
    2 Institute for Clinical Research and Health Policy Studies, Tufts Medical Center \\
    3 Department of Medicine, Boston University \\
    4 Social Science Research Institute, Duke University \\
    5 Department of Public Health, University of Southern Denmark \\
    6 Department of Mathematics and Statistics, Boston University \\}
  \maketitle
} \fi

\if0\blind
{
  \bigskip
  \bigskip
  \bigskip
  \begin{center}
    {\LARGE\bf Estimating the cognitive effects of statins from observational data using the survival-incorporated median: a summary measure for clinical outcomes in the presence of death}
\end{center}
  \medskip
} \fi

\bigskip
\begin{abstract}
The issue of ``truncation by death" commonly arises in clinical research: subjects may die before their follow-up assessment, resulting in undefined clinical outcomes. To address this issue, we focus on survival-incorporated quantiles---quantiles of a composite outcome combining death and clinical outcomes---to summarize the effect of treatment. Using inverse probability of treatment weighting (IPTW), we propose an estimator for survival-incorporated quantiles from observational data, applicable to settings of both point treatment and time-varying treatments. We establish consistency and asymptotic normality of the estimator under both the true and estimated propensity scores. While the variance properties of IPTW estimators for the mean have been studied, to our knowledge, this article is the first to show that the IPTW quantile estimator using the estimated propensity score yields lower asymptotic variance than the IPTW quantile estimator using the true propensity score. Extensive simulations show that survival-incorporated quantiles provide a simple and useful summary measure and confirm that using the estimated propensity score reduces the root mean square error. We apply our method to estimate the effect of statins on the change in cognitive function, incorporating death, using data from the Long Life Family Study (LLFS)---a multicenter observational study of 4953 older adults with familial longevity. Our results indicate no significant difference in cognitive decline between statin users and non-users with a similar age- and sex-distribution at baseline. This study not only contributes to understand the cognitive effects of statins but also provides insights into analyzing clinical outcomes in the presence of death.

\end{abstract}

\noindent%
\vfill

\newpage
\spacingset{1.9} 
\section{Introduction}
\label{s:intro}

\subsection{Background}

``Truncation by death" is a common challenge in clinical research, as subjects may die before the follow-up assessment, resulting in undefined clinical outcomes. This challenge is particularly prevalent in longitudinal studies of older adults, such as the Long Life Family Study (LLFS), an international multicenter observational study involving 4953 older adults with exceptional longevity \citep{wojczynski2022nia}. In the LLFS, death is common and closely related to many clinical outcomes, such as cognitive function \citep{arbeev2020composite, xiang2023signatures}. Furthermore, death can gradually alter the composition of the study population \citep{murphy2011treatment}. Therefore, simply treating death as censoring or missing data may lead to biased estimates and misleading conclusions \citep{colantuoni2018statistical, xiang2023survival}, which makes it essential to consider death carefully when analyzing such studies.

To address the issue of truncation by death,  we advocate summarizing the clinical benefit of treatment by combining death and the clinical outcome into a ranked composite outcome $\tilde{Y}$~\citep{lachin1999worst, chen2005treatment, lok2010long, wang2017inference}. Because this composite outcome integrates two outcomes on different scales (death and clinical outcomes), it is inappropriate to draw inference using the mean of the composite outcome. Instead, we assess the clinical benefit of treatment by comparing the distribution or quantiles of the composite outcome $\tilde{Y}$: survival-incorporated quantiles \citep{xiang2023survival}. In particular, one can focus on the survival-incorporated median, which provides as a simple and useful summary measure when the probability of death of the target population is less than 50\%, as conceptualized in \citet{xiang2023survival}.

This paper focuses on the estimation of survival-incorporated quantiles from observational data in the presence of death. To achieve this, we propose a weighted quantile estimator based on Inverse Probability of Treatment Weighting (IPTW) \citep{robins1986new, robins2000marginal, hernan2000marginal}. IPTW allows weighting composite outcomes, including death, to estimate the marginal quantiles of the potential outcomes $\tilde{Y}^{(a)}$ under the treatment of interest $a$.  We show that the proposed estimator performs well for point treatment settings with $a = 0$ and $a=1$. Furthermore, with carefully constructed weights, the estimator can be readily extended to settings with time-varying treatments, which are common in studies with long-term follow-up with subjects at risk of death, such as the LLFS.

In the causal inference literature, most studies focus on population means of potential outcomes. However, inference on quantiles \citep{hogan2004marginal, firpo2007efficient, zhang2012causal, sherwood2013weighted,  sun2021causal, cheng2024doubly} can be especially useful for the important problem of truncation by death. \cite{hogan2004marginal} studied the marginal structural quantile regression model and showed that their quantile estimator solves an unbiased estimating equation. However, due to the non-smooth nature of the quantile estimator and the complexity of nuisance parameters in the propensity score, consistency does not simply follow from an unbiased estimating equation. In this article, we prove both consistency and asymptotic normality for the proposed IPTW quantile estimator.

In particular, \citet{sun2021causal} showed that  the asymptotic limiting distribution for quantile treatment effect remains unchanged whether using the true or estimated propensity score, given that the estimated propensity score is uniformly consistent.  However, using a different proof strategy, we show that the asymptotic variance for the IPTW quantile estimator is smaller when using the estimated propensity score compared to the true propensity score. Our simulation results support this finding, as the root Mean Square Error (rMSE) is lower when using the estimated propensity score.  While similar variance properties of IPTW estimators for the mean have been discussed in previous research \citep{rotnitzky1995semiparametric, lunceford2004stratification, lok2024estimating},  properly estimating the variance received less attention in quantile estimation. To the best of our knowledge, this article is the first to establish the asymptotic variance of the IPTW quantile estimator using both the true and the estimated propensity scores, and to highlight their difference.

\subsection{Motivating clinical question}

Our clinical question of interest is the effect of statins on the change in cognitive function of LLFS participants. Statins, commonly prescribed to lower cholesterol and manage cardiovascular conditions, are used by nearly 30\% of adults 40 years and older in the United States \citep{schultz2018role}. Despite the widespread use of statins, their impact on cognitive function remains a debate \citep{ott2015statins, schultz2018role, adhikari2021association, ying2021impact, olmastroni2022statin}. Some studies indicate a potential risk of cognitive impairment from statins \citep{muldoon2000effects, alsehli2020cognitive}, while other studies suggest no significant risk or even a protective effect on cognitive function \citep{benito2010statins,  petek2023statins}. The complexity of the relationship between statins and cognitive function necessitates further research.

Many of the aforementioned studies address undefined clinical outcomes due to death inadequately. For example, \citet{alsehli2020cognitive} ``excluded all data sets containing missing values" and performed a survivors-only analysis. This survivors-only analysis is known to be affected by the ``healthy-survivors" effect and is subject to bias \citep{rothman2008modern}. \citet{petek2023statins} mention the considerable number of patients who dropped out, and they applied Inverse Probability of Censoring Weighting (IPCW) \citep{hernan2020causal} to handle death and missingness together. However, death leads to undefined outcomes rather than missing outcomes, so it is inappropriate to treat death the same as missingness; if death is treated as censoring, IPCW transfers the weight of dead patients to those alive. These two approaches, survivors-only analyses and IPCW censoring at death, are common in dealing with undefined outcomes due to death but may lead to biased conclusions \citep{robins1995analytic, zhang2003estimation,  xiang2023survival}. 

The objectives of this study are twofold: (i) to estimate the change in cognitive function of the LLFS participants on and off statins at baseline, while (ii) properly addressing the issue of truncation by death using survival-incorporated quantiles. Through this application, we aim to not only contribute to this clinical question regarding statins but also offer insights into analyzing clinical studies in the presence of death.

This paper is structured as follows. Section~\ref{s:notation} introduces the setting, the definition, and the assumptions for estimation of survival-incorporated quantiles in the LLFS. Section~\ref{s:estimation} describes the IPTW estimator of survival-incorporated quantiles in both point treatment settings and time-varying settings.  Section~\ref{s:statistical} provides statistical properties of the proposed estimators. Section~\ref{s:simulation} presents simulation studies to investigate the performance of the proposed estimators. Section~\ref{s:application} applies the survival-incorporated median to study the change in cognitive function of the LLFS participants. A discussion concludes the article in Section~\ref{s:discuss}.

\section{Setting, definition, and assumptions}
\label{s:notation}

\subsection{Setting and notation}

This article uses the following notation. Consider a study of $N$ participants with a baseline assessment ($k=0$), and $k = 1, ..., K + 1$ subsequent follow-up visits. $D_{k, i}$ is the indicator variable of the survival status at time $k$, with $D_{k, i} = 1$ if participant $i$ is dead and $D_{k, i} = 0$ if participant $i$ is alive at time $k$. $Y_i$ is the continuous outcome that is measured at time $K + 1$, the end of the study, in those alive ($D_{(K+1), i} = 0$); $\Tilde{Y}_i$ is the composite outcome that combines $Y_i$ and death. $L_{k, i}$ is a vector of covariates representing measured risk factors at time $k$, $k = 0, …, K$. $A_{k, i}$ is the treatment indicator at time $k$, with $A_{k, i} = 1$ if participant $i$ is on treatment at time $k$ and $A_{k, i} = 0$ if not. At time $k$, in those alive $(D_{k, i} = 0)$, a treatment decision $A_{k, i}$ is made after measuring $L_{k, i}$. $\bar{L}_{k, i} = (L_{0, i}, L_{1, i}, …, L_{k, i})$ is the covariate history from baseline to the $k$th visit, and similarly $\bar{A}_{k, i}=(A_{0, i},A_{1, i},…,A_{k, i})$. $Y^{(\bar{a}_{K})}_i$ is the potential outcome had participant $i$ received treatment regimen $\bar{a}_{K}=(a_{0}, a_{1}, …, a_{K})$. The observed data of participant $i$ at time $k$ consist of $(D_{k, i},L_{k, i},A_{k, i})$, $k=0,…,K$. At the last follow-up time $K+1$, we observe $D_{K+1, i}$ and $Y_i$ (if $D_{K+1, i} = 0$). If participant $i$ died between the $(m-1)$th visit and the $m$th visit, $m = 1,2,...,K+1$, then $(L_{k, i}, A_{k, i}, Y_i)$ for participant $i$ becomes undefined for $k >= m$. The full data are $(\bar{L}_{K,i}, \bar{A}_{K, i}, \bar{D}_{(K+1),i}, \Tilde{Y}_{i})$ for participants $i=1,\dots,N$.

\subsection{Definition of the survival-incorporated quantile}

The survival-incorporated quantile is a summary measure of the ranked composite outcome that combines death and a clinical outcome \citep{lok2010long, xiang2023survival}. The survival-incorporated $\tau$th quantile is defined when the probability of death is less than $\tau$, ensuring that such quantile corresponds to a clinical outcome rather than death. This allows for meaningful comparisons, as it is not informative to compare summary measures corresponding to ``death'' under both treatment and control groups. The survival-incorporated quantile is defined as follows:

\begin{definition}[survival-incorporated $\tau$th quantile] The threshold such that a (1-$\tau$) proportion  of the target population is alive with a better clinical outcome than this threshold, while a $\tau$ proportion either died or has a clinical outcome worse than this threshold.
\end{definition}

To estimate the survival-incorporated quantile, all outcomes need to be ranked and combined together into a composite outcome $\tilde{Y}_i$.  Considering death a worse state than being alive, we assign participants who died any value less than the worst clinical outcome. For example, in the LLFS, the DSST scores have a range of $[0, 93]$ with higher scores suggesting better cognitive function; we can assign those who died a value of -10, -100, or -1000 to rank all outcomes together.

Such assignment is conceptually different from imputing missing values. Imputing missing values, which typically depends on the missing data mechanism, aims to replace a missing value with an estimated value. However, the death of a participant is fully observed, and if a participant died, the participant's clinical outcome is not missing but rather undefined. Moreover, the value of the assigned clinical outcome, which has the lowest ranking in the ranked outcomes, is irrelevant to the value of the survival-incorporated $\tau$th quantile given that the probability of death is less than $\tau$. Such assignment for undefined outcomes facilitates the computation the survival-incorporated quantiles.

Mathematically, for the composite outcome $\tilde{Y}^{(\bar{a})}_{i}$ under treatment regimen $\bar{a}$ with distribution function $F_{\tilde{Y}^{(\bar{a})}}(y) = P(\tilde{Y}^{(\bar{a})}_{i} \leq y)$, the survival-incorporated $\tau$-th quantile is defined as 
\[
q^{(\bar{a})}_{\tau} = F_{\tilde{Y}^{(\bar{a})}}^{-1}(\tau) = \inf\{y: F_{\tilde{Y}^{(\bar{a})}}(y) \geq \tau\},
\]
In practice, it is often useful and convenient to focus on the survival-incorporated median:
\[
q^{(\bar{a})}_{0.5} = F_{\tilde{Y}^{(\bar{a})}}^{-1}(0.5) = \inf\{y: F_{\tilde{Y}^{(\bar{a})}}(y) \geq 0.5\}.
\]
We will consider the survival-incorporated median in simulation studies (Section 5) and the LLFS application (Section 6).

\subsection{Assumptions}
Estimating survival-incorporated quantiles from observational data relies on the following identifying assumptions:

\begin{assumption}[No Unmeasured Confounding]
\label{asp: no unmeasured confounding}
For $\bar{a}_K$ and all $\bar{l}_k,$ $k = 0, ..., K+1$, 
\newline
   $A_{k,i}  \ind (\bar{L}_{K+1, i}^{(\bar{a}_K)}, \tilde{Y}^{(\bar{a}_{K})}_{i})  \rvert \bar{A}_{k-1, i}=\bar{a}_{k-1}, \bar{L}_{k, i} = \bar{l}_{k}$. 
\end{assumption}

\begin{assumption}[Consistency]
\label{asp: consisitency}
For all $\bar{a}_k$ and all $k$, $k = 0, ..., K+1$, if $\bar{A}_{K, i} = \bar{a}_K$, $\tilde{Y}^{(\bar{a}_K)}_{i} = \tilde{Y}_{i}$, and if $\bar{A}_{k - 1, i} = \bar{a}_{k - 1}$,  $\bar{L}^{(\bar{a}_{k-1})}_{k, i} = \bar{L}_{k}$.
\end{assumption}

\begin{assumption}[Positivity]
\label{asp: positivity}
For all $(\bar{a}_{k}, \bar{l}_{k})$ and all $k = 0, ... , K$, there exists an $\varepsilon > 0$ such that $ P(A_{k,i} = a_k \rvert \bar{A}_{k - 1, i } =\bar{a}_{k - 1}, \bar{L}_{k, i} = \bar{l}_{k}) > \varepsilon$. 
\end{assumption}

Assumptions 1, 2, and 3 are common in the causal inference literature for identification and estimation of causal parameters from observational data  \citep{hernan2020causal}. {\it No Unmeasured Confounding} Assumption 1 requires that all confounders that could influence both the treatment and future potential outcomes are measured and accounted for. {\it Consistency} Assumption 2 requires that the potential outcome for a participant under the treatment they actually received is consistent with their observed outcome. {\it Positivity} Assumption 3 requires that participants with any characteristics have some chance of taking and of not taking the treatment at each time point.

Finally, in the application and in the proofs for the statistical properties of our method, the propensity score of the treatment is estimated through a logistic regression model, assuming such model is correctly specified: 

\begin{assumption} [Propensity score model]
\label{asp: PS model correct}
The logistic regression model for the propensity score, $logit(p_\theta (A_{k, i} = 1 | \bar{A}_{k-1, i}, \bar{L}_{k, i})) = \theta_0 + \theta^\top_1 g(\bar{A}_{k-1, i}, \bar{L}_{k, i})$,  is correctly specified, with $g$ a function of $\bar{A}_{k-1, i}$ and $\bar{L}_{k, i}$, and the true parameter $\theta^*$ in a compact space $\Theta \subset \mathbb{R}^{p}$.
\end{assumption}

\section{Estimation of survival-incorporated quantiles from observational data}
\label{s:estimation}

\subsection{Point treatment settings}
We first describe the IPTW estimator for survival-incorporated quantiles from observational data with a binary treatment $A_i=0$ or $A_i=1$: a “point treatment” setting. Figure 1 (a) depicts the Directed Acyclic Graph (DAG) for this setting. $L_i$ is a vector of baseline covariates. After receiving the treatment, participants may die before the follow-up assessment ($D_i=1$) or may survive with their clinical outcome $Y_i$ measured. 

\begin{figure}[h]
  \centerline{\includegraphics[width=0.8\textwidth]{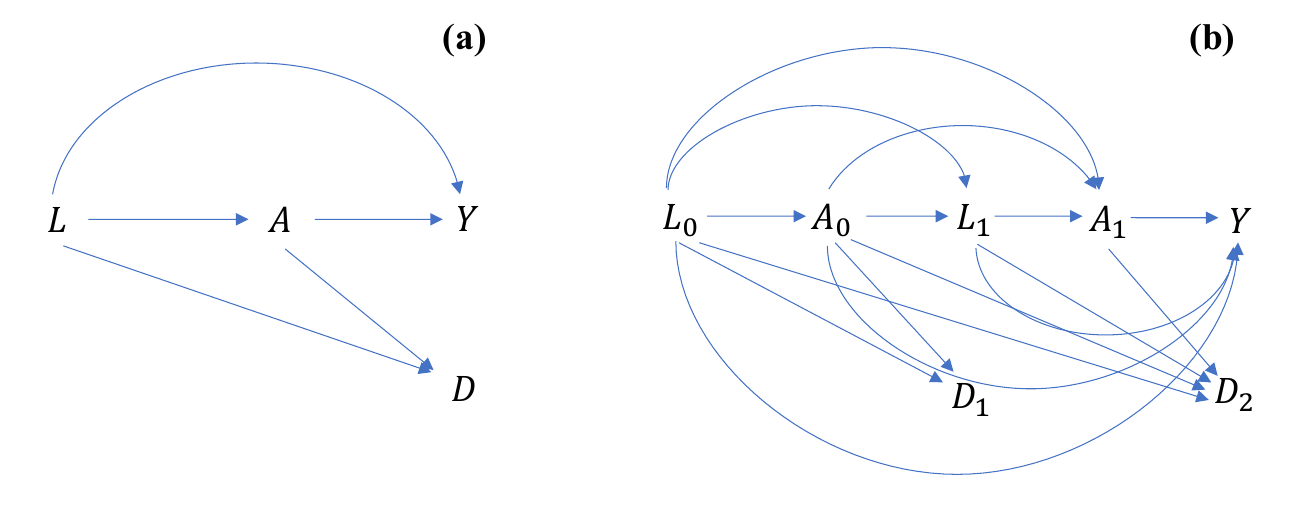}}
  \caption{DAG for (a) a point treatment setting and (b) a time-varying treatment setting with two post-baseline assessments.}
\label{f: DAG}
\end{figure}

IPTW is used to estimate the survival-incorporated $\tau$th quantile from observational data. First, as in Section 2.2, we assign those who died ($D_i=1$) a value less than the lowest possible value of $Y_i$.  Next, since the treatment is not randomized, we weight each outcome by the inverse of the participant’s probability of receiving their observed treatment conditional on the baseline covariates $L_i$, i.e., the inverse of the propensity score, $w_{a,i} = \mathbbm{1}_{A_i=a}/P(A_i=a|L_i)$. In observational studies, propensity scores are typically not known. We estimate the propensity scores $\hat{P}(A_i=a|L_i)$ assuming that a model for the probability of receiving the treatment is correctly specified, for example, a logistic regression model (Assumption 4).

Combining the weights with the quantile estimation procedure proposed by \citet{koenker1978regression}, the IPTW estimator for the survival-incorporated $\tau$th quantile under treatment $a$, 
$\hat{q}^{(a)}_{\tau}$, is
\begin{equation}  \label{eq: estimator}
\hat{q}^{(a)}_{\tau} = \argmin_{q} \frac{1}{N}\sum_{i=1}^{N} \hat{w}_{a, i} \cdot \rho_{\tau}(\tilde{Y_i} - q),
\end{equation}
where $\rho_\tau (x) = x(\tau - \mathbbm{1}_{x \leq 0})$ is the quantile loss function evaluated at $x$ (Koenker 2005), and $\hat{w}_{a,i}$ is the weight of the participant $i$ under treatment $a$, $a = 0$ or $a = 1$:
\[
\hat{w}_{0,i} = \frac{\mathbbm{1}_{A_i = 0}}{\hat{P}(A_i = 0|L_i)}, \; \hat{w}_{1,i} = \frac{\mathbbm{1}_{A_i = 1}}{\hat{P}(A_i = 1|L_i)}.
\]
The resulting estimating equations for $\hat{q}^{(a)}_{\tau}$ can be expressed as
\begin{align}
\label{eq: gradient function}
    \Psi_N(q) = \frac{1}{N}\sum^N_{i=1} \frac{\mathbbm{1}_{A_i = a}}{\hat{P}(A_i = a|L_i)}(\mathbbm{1}_{\tilde{Y}_i \leq q} - \tau) = 0.  
\end{align}
$\Psi_N(q)$ is the gradient function of the objective function in equation \eqref{eq: estimator}.

\subsection{Time-varying settings}
    Figure 1 (b) depicts the DAG of a time-varying setting with two post-baseline assessments ($K + 1 = 2$). In general, consider a study with $K+1$ follow-up assessments, where participants are at risk of death in each time interval.  The IPTW weights $\hat{w}_{\bar{a},i}$ need to account for death that may occur between each visit. Therefore, the IPTW weight of equation (1) is modified as follows:
    
    \begin{align*}
    & \hat{w}_{\bar{a},i} = 
    \begin{cases}
     \frac{\mathbbm{1}_{\bar{A}_{K,i} = \bar{a}_{K}}}{\prod^{K}_{k = 0}\hat{P}(A_{k,i} = a_k |\bar{A}_{k-1, i} =a_{k-1}, \bar{L}_{k, i})} \quad \begin{aligned}
                    & \text{if $ D_{(K+1),i} = 0$, i.e., } \\
                    & \text{participant $i$ survives throughout,}
                       \end{aligned} \\[15 pt]
     \frac{\mathbbm{1}_{\bar{A}_{M-1, i} = \bar{a}_{M-1}}}{\prod^{M-1}_{k = 0}\hat{P}(A_{k, i} = a_k|\bar{A}_{k-1, i} = a_{k-1}, \bar{L}_{k, i})} \, \; \; \begin{aligned}
                        & \text{if participant $i$ dies between the}\\
                        & \text{$(M - 1)$th visit and the $M$th visit.}
                        \end{aligned} 
    \end{cases}    
    \end{align*}

When participant $i$ dies between the $(M - 1)$th visit and the $M$th visit, their clinical outcome and covariates become undefined starting from time $M$. Such participant’s undefined covariates from time $M$ onwards are irrelevant. Hence,  the denominator of the weight is the participant’s probability of receiving the treatment history that they received, conditional on their covariate and treatment history before the time they died.

\section{Statistical properties of the estimated survival-incorporated quantile}
\label{s:statistical}

Theorem 1 below states that the survival-incorporated $\tau$th quantile of the outcome is identifiable from observational data. Theorem 1 generalizes Lemma 1 in \cite{firpo2007efficient} from point treatment settings to time-varying treatment settings with $K + 1$ follow-up times. Appendix A.1 provides the proof of Theorem 1.

\begin{theorem}[Identification of quantiles] 
\label{thm: identification}
Under Assumptions 1-4, the $\tau$th quantile of the composite outcome $\tilde{Y}$ under treatment regimen $\bar{a}_{K}, q^{(\bar{a}_{K})}_{\tau}$, can be expressed as an implicit function of the observed data:
\[
    E \Bigg(\frac{\mathbbm{1}_{\bar{A}_{M, i} = \bar{a}_{M}}}{\prod^{M}_{k = 0} P [A_{k, i} = a_k | \bar{A}_{k-1, i} = \bar{a}_{k - 1}, \bar{L}_{k, i}]} \cdot \big( \mathbbm{1}_{\tilde{Y}_{i} \leq q_{\tau}^{(\bar{a}_{K})}} - \tau \big)\Bigg) = 0,
\]
where M = K if participant $i$ survives throughout, or M is the last visit if participant $i$ died before the visit K+1.
\end{theorem}

Appendix A.2 includes additional regularity conditions for consistency and asymptotic normality. In particular, Condition  6 and 7 ensure that the density function $f_{\tilde{Y}}(y)$ is bounded away from zero near the target $\tau$th quantile for point treatment $a$ and time-varying treatment $\bar{a}$, respectively. Theorem 2 below states that the IPTW-estimator for the $\tau$th quantile is consistent for the true population survival-incorporated $\tau$th quantile. Appendix A.2 includes the proof of Theorem 2.

\begin{theorem}[Consistency]
\label{thm: consistency}
Under Assumptions 1-4 in Section 2 and regularity conditions 5-7 in Appendix A.2,
\begin{align*}
\hat{q}_{\tau}^{(a)} \overset{P}{\to} q_{\tau}^{(a)}, \\
\hat{q}_{\tau}^{(\bar{a})} \overset{P}{\to} q_{\tau}^{(\bar{a})} .
\end{align*}
\end{theorem}

For point treatment $a$, let $\tilde{q}^{(a)}_{\tau}$ denote the estimator for the $\tau$th quantile when the propensity score is known; Theorem 3 below states that $\tilde{q}^{(a)}_{\tau}$ is asymptotically normal. When the propensity score is estimated, Theorem 4 below states that $\hat{q}^{(a)}_{\tau}$ is asymptotically normal. Appendix A.3 and A.4 provide the proof of Theorem \ref{thm: asymptotic normality with known PS} and Theorem \ref{thm: asymptotic normality with estimated PS}, respectively. For time-varying treatment $\bar{a}$, asymptotic normality can be derived similarly.

\begin{theorem}[Asymptotic normality with known propensity score]
\label{thm: asymptotic normality with known PS}
Under Assumptions 1-4 in Section 2 and regularity conditions 5-7 in Appendix A.2,
\[
\sqrt{N} (\tilde{q}^{(a)}_{\tau} \to q^{(a)}_{\tau}) \overset{\mathcal{D}}{\to} N \bigg ( 0, \frac{\tilde{V}}{f^2_{\tilde{Y}^{(a)}}(q^{(a)}_{\tau})} \bigg),
\]
where
\[
\tilde{V} = E \bigg\{\bigg[\frac{\mathbbm{1}_{A_i = a}}{P(A_i = a |L_i)} \Big( \mathbbm{1}_{\tilde{Y}_i \leq q^{(a)}_{\tau}} - \tau \Big)\bigg]^{2}\bigg\}.
\]
\end{theorem}

\begin{theorem} [Asymptotic normality with estimated propensity score]
\label{thm: asymptotic normality with estimated PS}
Under Assumptions 1-4 in Section 2 and regularity conditions 5-7 in Appendix A.2,
\[
\sqrt{N} (\hat{q}^{(a)}_{\tau} \to q^{(a)}_{\tau}) \overset{\mathcal{D}}{\to} N \bigg ( 0, \frac{V}{f^2_{\tilde{Y}^{(a)}}(q^{(a)}_{\tau})} \bigg),
\]
where
\[
V = \tilde{V} -D^\top I\bigl(\theta^*\bigr)^{-1}D,
\]
with $\tilde{V}$ the variance from Theorem 3, $I\bigl(\theta^*\bigr)$ the partial Fisher information for $\theta$ from Assumption 4, $p_{\theta^*}(A_i = a |L_i)$ the true propensity score, and
\begin{equation*}D^\top=E\left( L_i^\top \mathbbm{1}_{A_i = a}   \left( 1 -  \frac{1}{p_{\theta^*}(A_i = a|L_i)} \right) \left(\mathbbm{1}_{\tilde{Y}_i \leq q^{(a)}_{\tau}} - \tau \right)\right). 
\end{equation*}
\end{theorem}

Since  $D^\top I\bigl(\theta^*\bigr)^{-1}D$ is non-negative,  $V < \tilde{V}$. Therefore, theorems \ref{thm: asymptotic normality with known PS} and \ref{thm: asymptotic normality with estimated PS} show that estimating quantiles based on the estimated propensity score is more efficient, as was also seen for estimating means \citep{robins1994estimation}.

\section{Simulation study}
\label{s:simulation}

This simulation study evaluates the performance of the proposed IPTW quantile estimator for the survival-incorporated median.

\subsection{Point treatment setting}
\begin{figure}
  \centerline{\includegraphics[width=5 in]{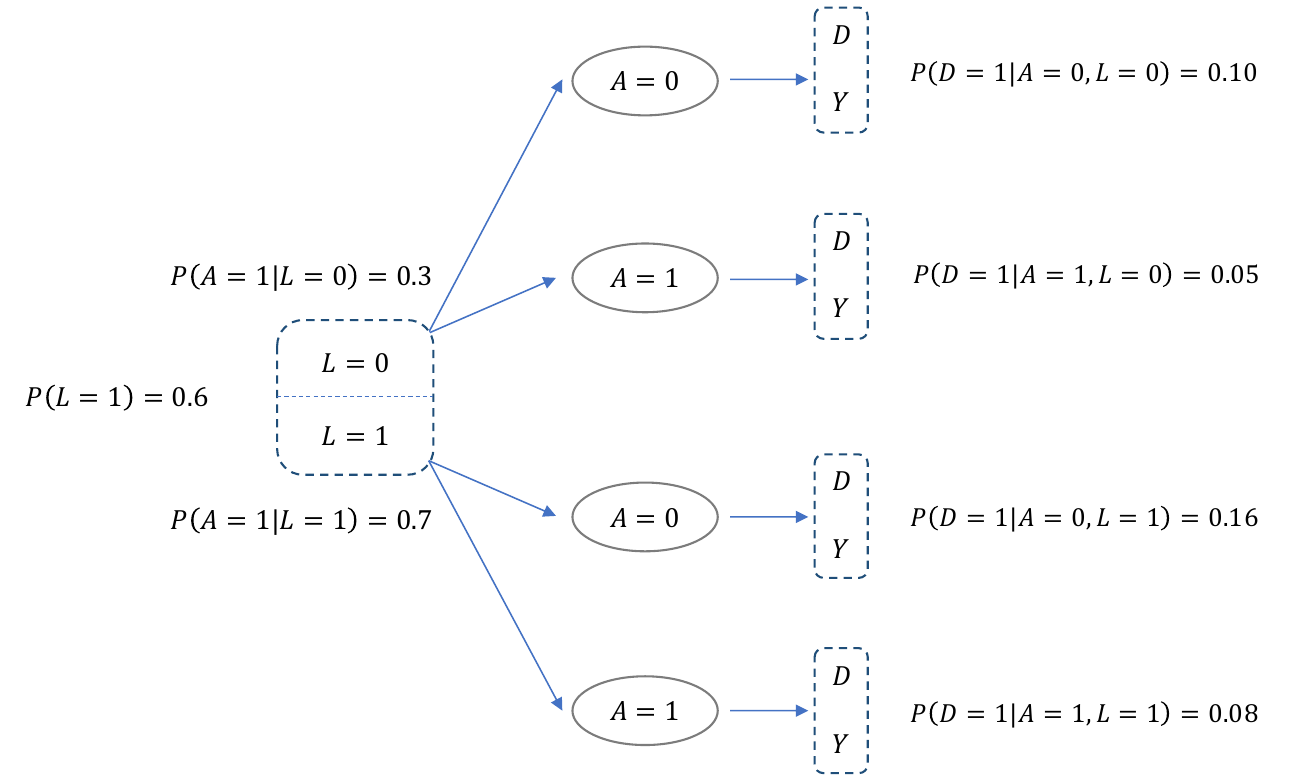}}
   \caption{Simulation scenario for a point treatment setting with observational data.}
\label{f:pointtrt}
\end{figure}

Figure \ref{f:pointtrt} depicts this simulation setting with a binary confounder $L_i$. In this simulation setting, treatment $a = 1$ improves survival but has a lower clinical outcome $Y_i$ in the survivors; we simulate the clinical outcome for those who survive by
\[
    Y_i = -0.9 A_i + 3 L_i + \varepsilon_i,  
\]
where $\varepsilon_i \sim N(0, 1)$.

For this simulation setting, Appendix B.2 provides a mathematical derivation of the true survival-incorporated median and the true median in the survivors under both $a=0$ and $a=1$. The true population survival-incorporated median is 1.449 under $a = 0$ and 0.915 under $a = 1$. The true population median in the survivors is 2.00 under $a = 0$ and 1.145 under $a = 1$. The survival-incorporated median shows a smaller difference between $a=0$ and $a=1$ compared to the median in the survivors.

\subsection{Time-varying setting}
This time-varying simulation setting has two follow-up times (Appendix Figure~B1). $L_{k, i}$ is a binary covariate for $k=0, 1$. At baseline, $A_{0, i}$ is assigned based on baseline covariate $L_{0, i}$. At the first follow-up time, the death status of participants is recorded, and if $D_{1, i} = 0$, covariate $L_{1, i}$ is observed. At the second follow-up time, the death status is again recorded, and if $D_{2, i} = 0$, the clinical outcome $Y_i$ is measured. We simulate the final clinical outcome for those who survive by
\[
    Y_i = 2L_{0,i} - 0.4A_{0,i} + 2.2L_{1,i} - 0.4A_{1,i} + \varepsilon_i,  
\]
where $\varepsilon_i \sim N(0, 1)$. The distribution of $L_{k, i}$, $A_{k, i}$, and $D_{k, i}$ all depend on the previous covariate and treatment history:
\begin{align*}
    & P(L_{0, i} = 1) = 0.6, \quad  P(A_{0, i} = 1 |L_{0, i} = 0) = 0.3, \quad P(A_{0, i} = 1 |L_{0, i} = 1) = 0.7, \\
    & \text{logit}(P(D_{1, i} = 1|L_{0, i}, A_{0, i}))  = -2.5 + 0.5 L_{0, i} - 0.6 A_{0, i}, \\
    & \text{logit}(P(L_{1, i} = 1|L_{0, i}, A_{0, i}, D_{1, i} = 0))  = -1 + 2 L_{0, i} - A_{0, i}, \\
    & \text{logit}(P(A_{1, i} = 1|A_{0, i}, L_{0, i}, D_{1, i} = 0, L_{1, i}))  = -2.5 + 0.8 L_{0, i} +3 A_{0, i} + L_{1, i}, \\
    & \text{logit}(P(D_{2, i} = 1|L_{0, i}, A_{0, i}, D_{1, i} = 0, L_{1, i}, A_{1, i})) = -3 + 0.3 L_{0, i} - 0.4 A_{0, i} + 0.5 L_{1, i}- 0.4 A_{1, i}.
\end{align*}
The coefficients in the above equations ensure that the probability of death under each treatment regimen is below 50\%. 

Two treatment regimens are of interest: $\bar{a} = (0,0)$ versus $\bar{a} = (1,1)$.  The true population survival-incorporated median is 1.726 under $\bar{a} = (0,0)$ and  0.751 under $\bar{a} = (1,1)$. The true population median in the survivors is 2.458 under $\bar{a} = (0,0)$ and 1.228 under $\bar{a} =(1,1)$~(Appendix B.2). Similar to the point treatment setting, the survival-incorporated median also shows a smaller difference between $a=0$ and $a=1$ compared to the median in the survivors.

\subsection{Simulation results}

We compare the proposed IPTW quantile estimator with an unweighted quantile estimator. For each simulation scenario, 2000 datasets were generated with the number of subjects $N = 500, 1500, 5000$. Table \ref{tab: sims point} shows the results for the point treatment setting and Table \ref{tab: sims time varying} shows the results for the time-varying setting. 

Appendix Table B1 summarizes the coverage probability of the bootstrap confidence intervals calculated  by the percentile method \citep{efron1992bootstrap}. Due to prolonged runtime, we consider two settings: $a = 1$ for point treatment and $\bar{a} = (1,1)$ for time-varying treatment. Each setting uses bootstrap sampling with 2000 replicates for 1000 simulated datasets with $N=1500, 5000$.

 \begin{table}[h]
\renewcommand{\arraystretch}{1.6}
\begin{center}
\small
\begin{tabular}{|c l c c c c c c c c|}
 \hline
\multirow{2}*{} & \multirow{2}*{} & \multirow{2}*{\makecell{True \\ $P$(death)}} 
& \multirow{2}*{Truth} 
& \multicolumn{2}{c}{\makecell{IPTW (true PS) \\ estimation} } & \multicolumn{2}{c}{\makecell{IPTW  (estimated PS) \\ estimation} } & \multicolumn{2}{c|}{\makecell{Unweighted \\ estimation}} \\
 \cline{5-10}     
                     &          &       &       & rMSE & Bias & rMSE & Bias  & rMSE & Bias \\
                     
\hline
\multirow{3}*{$a=0$} & $N$ = 500  & 0.136 & 1.449 & 0.308 & -0.003 & 0.275 & 0.006 & 0.975 & -0.961 \\ 
                     & $N$ = 1500 &       &       & 0.182 & -0.005 & 0.162 & -0.004 &0.975 & -0.970 \\ 
                     & $N$ = 5000 &       &       & 0.101 & -0.001 & 0.088 & 0.001 &0.971 & -0.970 \\ 
\hline
\multirow{3}*{$a=1$} & $N$ = 500  & 0.068 & 0.915 & 0.242 & -0.011 & 0.185 & -0.009 & 0.677 & 0.667 \\  
                     & $N$ = 1500 &       &       & 0.134 & -0.002 & 0.104 & -0.001 &0.673 & 0.670 \\ 
                     & $N$ = 5000 &       &       & 0.075 & -0.002 & 0.058 & -0.001 &0.670 & 0.669 \\
\hline
\end{tabular}
\end{center}
\caption{Simulation results for estimation of the survival-incorporated median in a point treatment setting. Truth: True survival-incorporated median. IPTW: Inverse Probability of Treatment Weighting. PS: propensity score. rMSE: root Mean Square Error.}
\label{tab: sims point}
\end{table}

\begin{table}[h]
\renewcommand{\arraystretch}{1.6}
\begin{center}
\small
\begin{tabular}{|c l c c c c c c c c|}
 \hline
\multirow{2}*{} & \multirow{2}*{} & \multirow{2}*{\makecell{True \\ $P$(death)}} 
& \multirow{2}*{Truth} 
& \multicolumn{2}{c}{\makecell{IPTW (true PS) \\ estimation} } & \multicolumn{2}{c}{\makecell{IPTW  (estimated PS) \\ estimation} } & \multicolumn{2}{c|}{\makecell{Unweighted \\ estimation}} \\
 \cline{5-10}     
                     &          &       &       & rMSE & Bias & rMSE & Bias & rMSE & Bias   \\
                     
\hline
\multirow{3}*{$\bar{a}=(0, 0)$} & $N$ = 500  & 0.170 & 1.726 & 0.359 & -0.004 & 0.325 & -0.004 & 0.776 & -0.743 \\ 
                                & $N$ = 1500 &       &       & 0.210 & -0.006 &  0.194 & -0.004 & 0.762 & -0.752 \\ 
                                & $N$ = 5000 &       &       & 0.114 & -0.004 &  0.103 & -0.002 & 0.756 & -0.753 \\ 
\hline
\multirow{3}*{$\bar{a}=(1, 1)$} & $N$ = 500  & 0.089 & 0.751 & 0.253 & 0.001  &  0.214 & -0.003 & 1.093 & 1.077 \\ 
                                & $N$ = 1500 &       &       & 0.141 & 0.003  & 0.121 & -0.002 & 1.084 & 1.078  \\ 
                                & $N$ = 5000 &       &       & 0.077 & 0.001  &  0.066 & 0.001 & 1.081 & 1.079 \\ 
\hline
\end{tabular}
\end{center}
\caption{Simulation results for estimation of the survival-incorporated median in a time-varying setting. Truth: True survival-incorporated median. IPTW: Inverse Probability of Treatment Weighting. PS: propensity score. rMSE: root Mean Square Error.}
\label{tab: sims time varying}
\end{table}

The simulation results show: (1) The unweighted quantile estimator is substantially biased, but the IPTW quantile estimator has a very small bias. (2) Both rMSE (root Mean Square Error) and bias decrease as the number of participants increases. (3) For every setting, the estimator for the survival-incorporated median based on the known propensity score  has a greater rMSE than the estimator based on the estimated propensity score. This aligns with the theory in Section \ref{s:statistical} that the estimator based on the estimated propensity score is more efficient. (4)~The bootstrap 95\% confidence intervals all have a coverage probability of approximately 95\%, indicating that the bootstrap is a valid tool for statistical inference of the proposed IPTW quantile estimator.

\section{Application: cognitive change in older adults on and off statins}
\label{s:application}

\subsection{LLFS study}

The LLFS participants were enrolled at three American field centers in Boston, Pittsburgh, and New York, as well as a Danish field center. The first in-person visit took place between 2006 and 2009, and the second in-person visit took place 8 years later using the same protocols. The cognitive function of the LLFS participants was assessed at these two in-person assessments 8 years apart by the Digit Symbol Substitution Test (DSST), a well-known neuropsychological test for measuring cognitive function \citep{wechsler1981wechsler}. Over the 8 years of follow-up, statins use was measured at baseline, at year 3, and at year 6. This leads to the following timeline: $k = 0$ (baseline), $k = 1$ (year 3), $k = 2$ (year 6), and $k = 3$ (year 8). A participant might die in three time intervals: $(0, 3], (3, 6]$, and $(6, 8]$. In the LLFS, the confounders were only measured at baseline, and the vector of baseline confounders $L_{0, i}$ includes: age at baseline, gender, education, smoking, total cholesterol level, low-density lipoproteins, high-density lipoproteins, and adjusted Framingham risk score (see Appendix C.1). The full data include the participants' baseline confounders $L_{0, i}$, treatment history $\bar{A}_{2, i}$, death status $D_{k, i}, k = 1,2,3$, and DSST scores $Y_i$ at $k = 3$ in those alive. 

\begin{table}
\centering
\centerline{\includegraphics[width = 3.5 in]{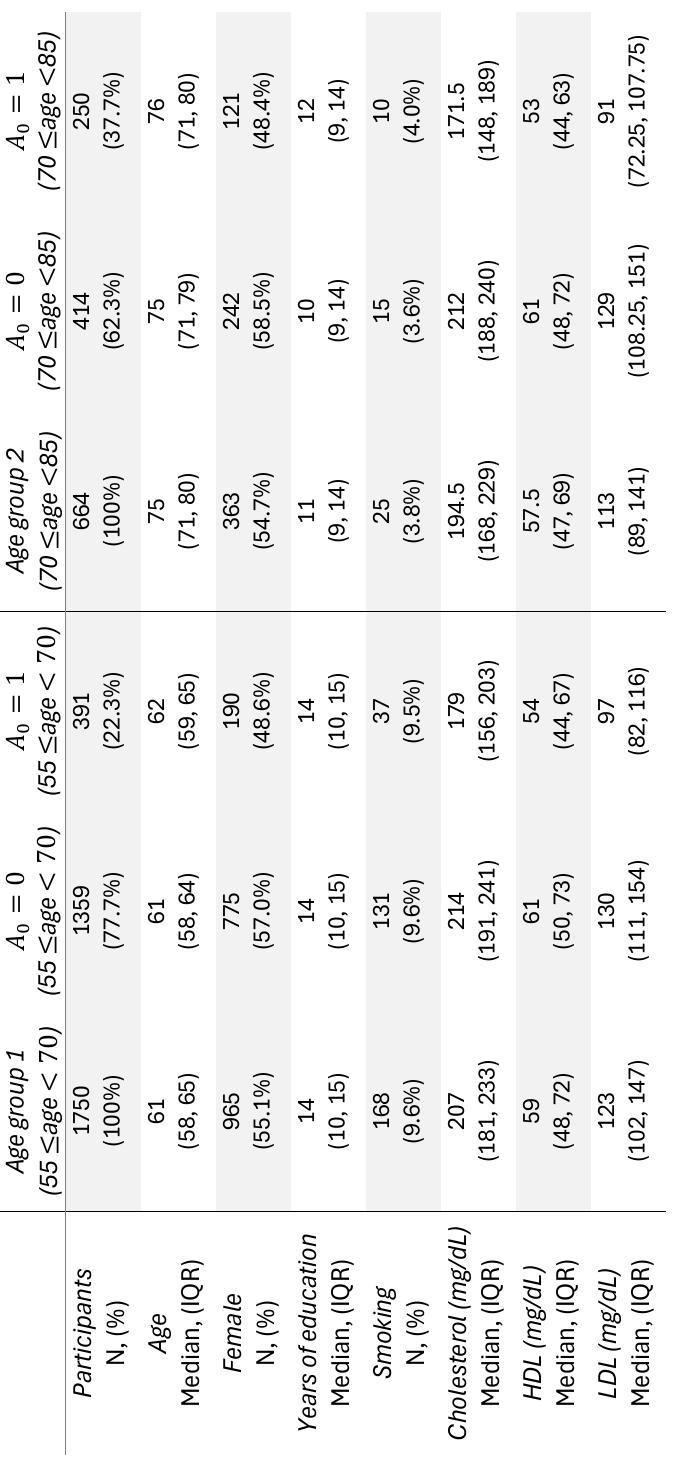}}

\caption{Baseline characteristics of the Long Life Family Study participants in two age groups; N, number; IQR, interquartile range; HDL, high-density lipoprotein; LDL, low-density lipoprotein; mg/dL, milligrams per deciliter; $A_{0} = 0$, participants off statins at baseline; $A_{0} = 1$, participants on statins at baseline.}
\label{tab: LLFS baseline}
\end{table}

We focus on two sub-populations of participants who had a baseline DSST assessment: (1) 1750 participants with age-at-enrollment between 55 and 69 ($55.0 \leq\text{age} < 70.0$) and (2) 664 participants with age-at-enrollment between 70 and 84 ($70.0 \leq \text{age} < 85.0$). Table \ref{tab: LLFS baseline} shows the baseline characteristics of the participants in the two age-based sub-populations.  Table \ref{tab: death status LLFS} shows the number of the participants on and off statins at baseline and the crude percentage of participants who died before the year-8 DSST.

\begin{table}[h]
\renewcommand{\arraystretch}{1.2}
\begin{center}

\begin{tabular}{|c c c c|}
\hline
 & & \makecell{\#Participants  } & \makecell{Crude \% of participants who died} \\
\cline{2-4}
\multirow{2}{*}{age 55 - 69}  & $A_0=0$  & 1359  & 4.1\%  \\ 
                              & $A_0=1$  & 391   & 4.6\%  \\ 

\hline
\multirow{2}{*}{age 70 - 84}  & $A_0=0$  & 414  & 17.6\%  \\ 
                              & $A_0=1$  & 250  & 19.2\%  \\ 

\hline
\end{tabular}

\caption{Death status of the Long Life Family Study participants in our analysis. $A_0=0$: participants off statins at baseline. $A_0=1$: participants on statins at baseline. In each age group, the crude probability of death is calculated by the number of deaths before the year-8 test divided by the number of participants.}
\label{tab: death status LLFS}
\end{center}
\end{table}

\subsection{Estimation}
We apply our method to estimate the survival-incorporated cognitive change in the LLFS participants on and off statins. This application has three main challenges: (1) some DSST scores are undefined due to death, (2) the LLFS is an observational study, and (3) some DSST scores are missing. In this application, we address (1) using the survival-incorporated median, (2) using Inverse Probability of Treatment Weighting (IPTW), and (3) using Inverse Probability of Censoring Weighting (IPCW).

\begin{figure}[h]
  \centerline{\includegraphics[width=2.4 in]{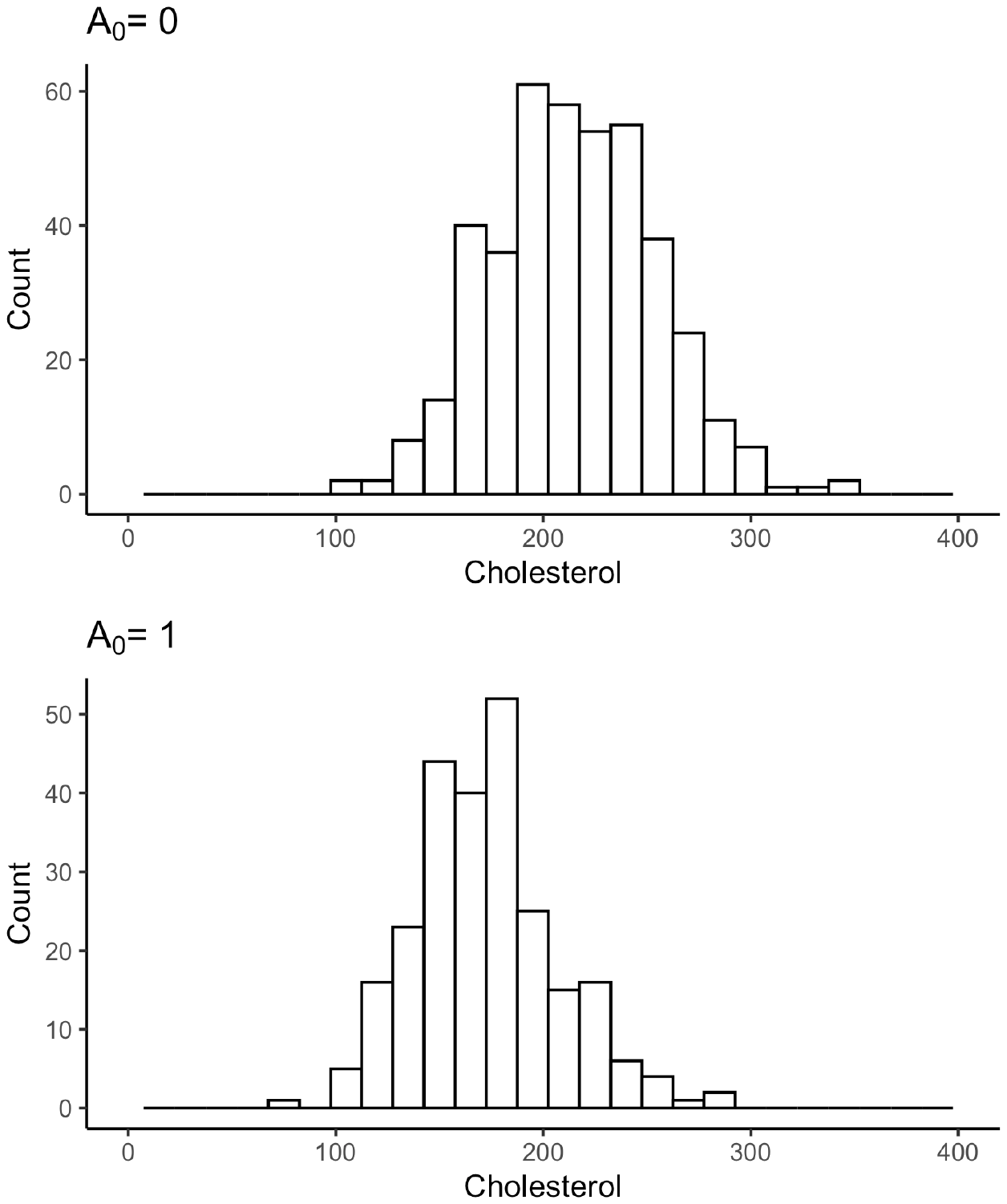}}
  \caption{Baseline cholesterol levels in those off statins at baseline $(A_{0, i} = 0)$ and on statins at baseline $(A_{0, i} = 1)$ in the LLFS participants with age in [70, 85).}
\label{fig:cholestero levels}
\end{figure}

The LLFS does not collect treatment information before baseline. Figure \ref{fig:cholestero levels} shows the distributions of total cholesterol in those off statins $(A_{0, i} = 0)$ and on statins $(A_{0, i} = 1)$ at baseline. The $A_{0, i}=0$ group has higher cholesterol levels compared to the $A_{0, i}=1$ group, suggesting participants who are on statins at baseline might have been using statins for some time. Therefore, the IPTW weights will not consider cholesterol levels at baseline as a covariate. However, given the correlation of cognitive function with age and sex in older adults \citep{harada2013normal, murman2015impact, levine2021sex, leshchyk2023mosaic}, the IPTW weights will consider age and sex as covariates.

Therefore, the objective of this application is to compare the survival-incorporated median cognitive change of the DSST score between baseline and 8 years in participants with a similar age- and sex-distribution off/on statins, had they remained off/on statins throughout. We use $a=0$ and $a=1$ to represent those two groups:
\begin{itemize}
    \item Group $a=0$: participants off statins at baseline, had they remained off statins throughout. 
    \item Group $a=1$: participants on statins at baseline, had they remained on statins throughout. 
\end{itemize}

We apply IPTW to account for the difference in the age distribution at baseline and the sex ratio between $A_{0, i}=0$ and $A_{0, i}=1$.  To compute the IPTW weights $\hat{w}_{i}^A$, the propensity score $\hat{p}(A_{0, i} | L^{age}_{i}, L^{sex}_{i})$, is estimated using a logistic regression model with age at baseline and sex as predictors:
\begin{align*}
    \text{logit}(p(A_{0, i} = 1 | L^{age}_{i}, L^{sex}_{i})) = \beta_0 + \beta_1 L^{age}_{i} + \beta_2   L^{sex}_{i}.
\end{align*}
After applying IPTW, the distributions of age at baseline and sex are comparable between the two groups (see Appendix Table C2 of weighted baseline characteristics).

We apply IPCW \citep{robins1994estimation} to account for the censored DSST scores due to the following reasons:
\begin{enumerate}
    \item Participant deviated from their initial treatment (participant off statins at baseline starting statins later, and participant on statins at baseline stopping statins later). We assume Missing At Random (MAR,  \citep{rubin1976inference}) and apply time-dependent IPCW.  
    \item Participant was absent for the year-8 DSST visit. We denote this as $C_{missing, i} = 1$ (versus $C_{missing, i} = 0$). We assume MAR and apply time-independent IPCW.  
    \item Participant attended the year-8 DSST assessment, but the result was invalid. We denote this as $C_{invalid, i} = 1$ (versus $C_{invalid, i} = 0$). We assume Missing Completely At Random (MCAR, \citep{rubin1976inference}) and apply IPCW without including covariates.
\end{enumerate}

The IPCW from (1) and (2) above are only conditioning on the baseline covariates $L_{0, i}$ listed in Table \ref{tab: LLFS baseline}, since the covariates are only measured at baseline. Statin usage is monitored at baseline, the first visit at year 3, and the second visit at year 6. All IPCW models are fitted separately for $A_{0, i} = 0$ and $A_{0, i} = 1$. For example, for those who are on statins at baseline, $A_{0, i} = 1$, the IPCW weights $\hat{w}_{A_0 = 1, i}^{C}$ are:
\begin{itemize}
    \item If participant $i$ was alive with a valid non-missing year-8 DSST score:
        \begin{align*}
     & \hat{w}_{A_0 = 1, i}^{C} =  \frac{\mathbbm{1}_{on\ statins\ through\ visit\ 2, i}}{\prod^{2}_{k=1} P(A_{k, i} = 1 | A_{k-1, i} = 1,, L_{0, i} , A_{0, i} = 1, D_{k, i} = 0)} \times   \\ 
     & \frac{1}{ P(C_{missing, i} = 0 | on\ statins\ through\ visit\ 2_{i}, L_{0, i} , A_{0, i} = 1, D_{2, i} = 0) } \times \\
     & \frac{1}{P(C_{invalid, i} = 0 | C_{missing, i} = 0, on\ statins\ through\ visit\ 2_{i}, A_{0, i} = 1, D_{2, i} = 0) }. 
    \end{align*}  
    The first factor is used to account for deviations from the initial regimen, the second factor is used to account for missing year-8 test scores, and the third factor is used to account for invalid year-8 test scores. $P(A_{k, i} = 1  | A_{k-1, i} = 1, L_{0, i} , A_{0, i} = 1, D_{k, i} = 0)$ is estimated by two logistic regression models: one for $k=1$ and one for $k=2$. 
$P(C_{missing, i} = 0 | on\ statins\ through\ visit\ 2_{i}, L_{0, i} , A_{0, i} = 1, D_{2, i} = 0)$ is estimated by a logistic regression model. 
$P(C_{invalid, i} = 0 | C_{missing, i} = 0, on\ statins\ through\ visit\ 2_{i}, A_{0, i} = 1, D_{2, i} = 0)$ is estimated by its empirical fraction.

    \item If participant $i$ died between the $(M-1)$th visit and the  $M$th visit, $M = 1, 2, 3$:
    \begin{align*}
        \hat{w}_{A_0 = 1, i}^{C} = \frac{\mathbbm{1}_{on\ statins\ until\ death, i}} {\prod^{M-1}_{k=1} P(A_{k, i} = 1 | A_{k-1, i} = 1, L_{0, i} , A_{0, i} = 1, D_{k, i} = 0)}.
    \end{align*}    
    If participant $i$ died between baseline and the first visit, their weight is 1.
    \item If participant $i$ stopped statins before the year-8 DSST, missed the year-8 DSST, or had an invalid year-8 DSST:
    \begin{align*}
         \hat{w}_{A_0 = 1, i}^{C} =  0.
    \end{align*}
\end{itemize}

The total weight for those on statins at baseline is the product of the IPTW weight and the IPCW weight, $\hat{w}_{i}^{A} \cdot \hat{w}^{C}_{A_0= 1, i}$. Similarly, for those off statins at baseline,  the total weight is $\hat{w}_{i}^{A} \cdot \hat{w}^{C}_{A_0= 0, i}$. Equation \eqref{eq: estimator} is then used to estimate the survival-incorporated median cognitive change in the DSST scores between baseline and 8 years.

\subsection{Results}

Table \ref{tab: LLFS sim res application} shows the estimated probabilities of death and the estimated survival-incorporated median cognitive change in age- and sex-comparable participants on statins $(a=1)$ and off statins $(a=0)$, had they continued their initial treatment throughout. In the age group 55-69, there is no difference in estimated survival-incorporated median cognitive change between group $a=1$ and group $a=0$. In the age group 70-84, compared to group $a=0$,  group $a=1$ is estimated to have one score less of survival-incorporated median cognitive decline. Considering that the DSST scores range from 0 to 100, the estimated differences of the survival-incorporated median cognitive change are relatively small. Moreover, the 95\% confidence interval of the difference (calculated by the bootstrap percentile method \citep{efron1992bootstrap}) is $[-1, 1]$ for the age group 55-69 and $[-2, 4]$ for the age group 70-84. These confidence intervals are relatively narrow and include zero. Therefore, our results indicate no statistically or clinically significant difference of cognitive change incorporating death between age- and sex-comparable participants on statins $(a=1)$ and off statins $(a=0)$, had they continued their initial treatment throughout.

\begin{table}[H]
\renewcommand{\arraystretch}{1.6}
\begin{center}
\begin{tabular}{|p{1.6cm} c c c|}
\hline
 & & \makecell{Estimated $\%$ of death \\ (95\% CI) } & \makecell{Survival-incorporated median \\ cognitive change  (95\% CI)} \\
\cline{2-4}
\multirow{3}{*}{age 55 - 69}  & $a=0$               & 4.5\%  (3.4\%, 5.7\%)   &-4 [-4, -3]\\ 
                              & $a=1$             & 3.2\%  (1.3\%, 5.4\%)   & -4 [-5, -3] \\ 
                              & $a=1$ - $a=0$     & -1.3\% (-3.5\%, 1.2\%) & 0 [-1, 1]\\
\hline
\multirow{3}{*}{age 70 - 84}  & $a=0$            & 18.2\% (14.1\%, 22.6\%)  & -8 [-9, -6]  \\ 
                              & $a=1$            & 14.8\%  (9.9\%, 20.5\%) & -7 [-9, -5] \\ 
                              &  $a=1$ - $a=0$    & -3.4\%   (-9.8\%, 3.5\%)  & 1  [-2, 4]\\

\hline
\end{tabular}
\caption{Probabilities of death and the survival-incorporated median cognitive change of the DSST scores between 8 years and baseline. Results are estimated with Inverse Probability of Treatment Weighting (IPTW) and Inverse Probability of Censoring Weighting (IPCW), ensuring participants are age- and sex-comparable. $a = 0$: participants off statins at baseline, had they remained off statins throughout. $a = 1$: participants on statins at baseline, had they remained on statins throughout. 95\% confidence intervals (CIs) are calculated by the bootstrap percentile method.  }
\label{tab: LLFS sim res application}
\end{center}
\end{table}

To compare with the median in the survivors, Appendix Table C3 presents the estimated median cognitive change in the survivors.  The estimated difference between the group $a=1$ and $a=0$ is -1 for the age group 55-69 and 0 for the age group 70-84, and the confidence intervals include zero. These results do not have a causal interpretation due to the selection bias from only including the survivors, who are inherently different between $a=1$ and $a=0$.

\section{Discussion}
\label{s:discuss}
To address the issue of ``truncation by death" in observational data, we propose an IPTW quantile estimator to estimate survival-incorporated quantiles to assess the clinical benefit of treatment. We prove consistency and asymptotic normality of the proposed estimator, and we demonstrate its performance through simulations. In the LLFS application, the estimated survival-incorporated median differences between age- and sex-comparable participants off and on statins are small, and the 95\% CIs are relatively narrow and include zero. Therefore, the LLFS application reveals no statistically or clinically significant differences of cognitive change between age- and sex-comparable participants on and off statins, incorporating death.

To estimate the survival-incorporated median, the IPTW estimator from Equation (1) may not be the only option. Estimation of a weighted quantile has the advantage of not being complex, and most standard statistical software supports such estimation, for example, the ``weighted\_quantile" function in the R package ``MetricsWeighted", as well as PROC MEANS with a WEIGHT statement in SAS.

In the simulation studies and the LLFS application, we compare the survival-incorporated median with the median in the survivors. \citet{xiang2023survival} compared the survival-incorporated median with the Survivor Average Causal Effect (SACE), the effect of treatment in those who survive regardless of the treatment option chosen. Identifying the SACE requires strong assumptions. In contrast, the survival-incorporated median relies on fewer assumptions and is simpler to estimate, making it a practical tool to summarize the clinical benefit of treatments on clinical outcomes in the presence of death \citep{xiang2023survival}.

In conclusion, our contributions are twofold. First, our findings of no significant difference in cognitive change incorporating death between age- and sex-comparable statin users and non-users, can inform strategies for statin prescription. This is particularly relevant for older adults, as statins are widely prescribed for cholesterol management. Second, the LLFS application demonstrates that the survival-incorporated median can serve as a practically useful summary measure of clinical outcomes in studies with mortality.

\bibliographystyle{plainnat}
\bibliography{Bibliography.bib}



\renewcommand{\thesection}{\Alph{section}}
\setcounter{section}{0}

\counterwithin*{equation}{section}
\renewcommand\theequation{\thesection\arabic{equation}}

\setcounter{table}{0}
\renewcommand{\thetable}{\Alph{section}\arabic{table}}

\setcounter{figure}{0}
\renewcommand{\thefigure}{\Alph{section}\arabic{figure}}

\newpage
\setcounter{page}{1} 
\begin{center}
    {\large\bf APPENDIX}
\end{center}

\section{Proofs}

\subsection{Proof of Theorem \ref{thm: identification}: identification of quantiles }
\label{a.sec:proof.thm1}
\text{  }

We first show that on the event $\bar{A}_{M, i} = \bar{a}_{M}$ with time $M \leq K$,
\begin{align}
    P [\bar{A}_{M,i} = \bar{a}_M | \bar{L}^{(\bar{a}_{K})}_{K, i} = \Bar{l}_{K}, \tilde{Y}_i^{(\bar{a}_{K})}] = \prod^{M}_{k = 0} P [A_{k, i} = a_k | \bar{A}_{k-1, i} = \bar{a}_{k - 1},\bar{L}_{k, i} = \Bar{l}_{k}], \label{eq:B product of propensity score}
\end{align}
as follows:
\begin{align*}
& P [A_{0, i} = a_0, A_{1, i} = a_1 ,..., A_{M, i} =a_{M} | \bar{L}^{(\bar{a}_{K})}_{K, i} = \Bar{l}_{K}, \tilde{Y}_i^{(\bar{a}_{K})}] \\
 & = P [A_{0, i} = a_0 | \bar{L}^{(\bar{a}_{K})}_{K, i} = \Bar{l}_{K}, \tilde{Y}^{(\bar{a}_{K})}_{i}] \cdot P [A_{1, i} = a_1 | A_{0, i} = a_0, \bar{L}^{(\bar{a}_{K})}_{K, i} = \Bar{l}_{K}, \tilde{Y}^{(\bar{a}_{K})}_{i}] \\ 
 & \qquad \cdots P[A_{M, i} = a_{M} | \bar{A}_{M-1, i} = \bar{a}_{M-1}, \bar{L}^{(\bar{a}_{K})}_{K, i} = \Bar{l}_{K}, \tilde{Y}_i^{(\bar{a}_{K})} ] \\
& = P [A_{0, i} = a_0 | L_{0, i} = l_0, \bar{L}^{(\bar{a}_{K})}_{K, i} = \Bar{l}_{K}, \tilde{Y}_i^{(\bar{a}_{K})}] \cdot P [A_{1, i} = a_1 | A_{0, i} = a_0, \Bar{L}_{1, i} = \Bar{l}_{1, i}, \bar{L}^{(\bar{a}_{K})}_{K, i} = \Bar{l}_{K}, \tilde{Y}^{(\bar{a}_{K})}_{i}] \\ 
 & \qquad  \cdots P[A_{M, i} = a_{M} | \bar{A}_{M-1, i} = \bar{a}_{M-1}, \bar{L}_{M, i} = \Bar{l}_{M}, \bar{L}^{(\bar{a}_{K})}_{K, i} = \Bar{l}_{K}, \tilde{Y}^{(\bar{a}_{K})}_{i}] \\
&  = P [A_{0, i} = a_0 | L_{0, i} = l_0] \cdot P [A_{1, i} = a_1 | A_{0, i} = a_0, \Bar{L}_{1, i} = \Bar{l}_1]   \\
 & \qquad \dotsm P[A_{M, i} = a_{M} | \bar{A}_{M-1, i} = \bar{a}_{M-1}, \bar{L}_{M, i} = \Bar{l}_{M} ].   
\end{align*}
The second equality uses Consistency (Assumption 2), and the third equality uses No Unmeasured Confounding (Assumption 1).

Equation \eqref{eq:B product of propensity score} is used to prove Theorem 1: 
\begin{align*}
&  E \Bigg(\frac{\mathbbm{1}_{\bar{A}_{M, i} = \bar{a}_{M}}}{\prod^{M}_{k = 0} P [A_{k, i} = a_k | \bar{A}_{k-1, i} = \bar{a}_{k - 1}, \bar{L}_{K, i}]} \cdot \big( \mathbbm{1}_{\tilde{Y}_i\leq q_{\tau}^{(\bar{a}_{K})}} - \tau \big)\Bigg) \\
= & E \Bigg(\frac{\mathbbm{1}_{\bar{A}_{M, i} = \bar{a}_{M}}}{\prod^{M}_{k = 0} P [A_{k, i} = a_k | \bar{A}_{k-1, i} = \bar{a}_{k - 1}, \bar{L}_{K, i}]} \cdot \big( \mathbbm{1}_{\tilde{Y}_i^{(\bar{a}_{K})} \leq q_{\tau}^{(\bar{a}_{K})}} - \tau \big)\Bigg) \\
= & E \Bigg(E \Bigg[\frac{\mathbbm{1}_{\bar{A}_{M, i} = \bar{a}_{M}}}{\prod^{M}_{k = 0} P [A_{k, i} = a_k | \bar{A}_{k-1, i} = \bar{a}_{k - 1}, \bar{L}_{K, i}]} \cdot \big( \mathbbm{1}_{\tilde{Y}_i^{(\bar{a}_{K})} \leq q_{\tau}^{(\bar{a}_{K})}} - \tau \big) \Big \rvert \bar{L}^{(\bar{a}_{K})}_{K, i}, \tilde{Y}_i^{(\bar{a}_{K})} \Bigg]\Bigg) \\ 
= & E \Bigg(\frac{ E [ \mathbbm{1}_{\bar{A}_{M, i} = \bar{a}_{M}} \rvert \bar{L}^{(\bar{a}_{K})}_{K, i}, \tilde{Y}_i^{(\bar{a}_{K})} ] }{\prod^{M}_{k = 0} P [A_{k, i} = a_k | \bar{A}_{k-1, i} = \bar{a}_{k - 1}, \bar{L}_{K, i}]} \cdot \big( \mathbbm{1}_{\tilde{Y}_i^{(\bar{a}_{K})}\leq q_{\tau}^{(\bar{a}_{K})}} - \tau \big)  \Bigg) \\
= & E \Bigg(\frac{P [\bar{A}_{M, i} = \bar{a}_M | \bar{L}^{(\bar{a}_{K})}_{K, i}, \tilde{Y}_i^{(\bar{a}_{K})}]}{\prod^{M}_{k = 0} P [A_{k, i} = a_k | \bar{A}_{k-1, i} = \bar{a}_{k - 1}, \bar{L}_{K, i}]} \Big( \mathbbm{1}_{\tilde{Y}_i^{(\bar{a}_{K})} \leq q_{\tau}^{(\bar{a}_{K})}} - \tau \Big)  \Bigg) \\
= & E \Bigg(\frac{\prod^{M}_{k = 0} P [A_{k, i} = a_k | \bar{A}_{k-1, i} = \bar{a}_{k - 1}, \bar{L}_{K, i}]}{\prod^{M}_{k = 0} P [A_{k, i} = a_k | \bar{A}_{k-1, i} = \bar{a}_{k - 1}, \bar{L}_{K, i}]} \Big( \mathbbm{1}_{\tilde{Y}_i^{(\bar{a}_{K})} \leq q_{\tau}^{(\bar{a}_{K})}} - \tau \Big)  \Bigg)  \\
= & E \big( \mathbbm{1}_{\tilde{Y}_i^{(\bar{a}_{K})} \leq q_{\tau}^{(\bar{a}_{K})}} - \tau \big) \\
= & F_{\tilde{Y}_i^{(\bar{a}_{K})}} \big(q_{\tau}^{(\bar{a}_{K})}\big) - \tau  = 0.
\end{align*}
The first equality uses Consistency (Assumption \ref{asp: consisitency}). The second equality uses the Law of Iterated Expectations. The fifth equality uses equation \eqref{eq:B product of propensity score}. The final equality uses the definition of the quantile $q_{\tau}^{(\Bar{a}_{K})}$.

\hfill $\square$

\subsection{Proof of Theorem \ref{thm: consistency}: consistency}
\text{ }

The proofs of Theorems 2, 3, and 4 require additional conditions.

\begin{condition}
\label{asp: compact L}
There exists a compact set $G \subset \mathbb{R}^{p}$ such that $P(L_i \in G) = 1$.
\end{condition}

\begin{condition}
\label{asp: continuous f(y) point trt}
For point treatment $a$, $\tilde{Y}^{(a)}_i$ has a density function $f_{\tilde{Y}^{(a)}}(y)$ that is bounded away from 0 in an open neighborhood of $q_{\tau}^{(a)}$.
\end{condition}

\begin{condition}
\label{asp: continuous f(y) tv trt}
For time-varying treatment $\bar{a}_{K}$, $\tilde{Y}^{(\bar{a}_{K})}_i$ has a density function $f_{\tilde{Y}^{(\bar{a}_{K})}}(y)$ that is bounded away from 0 in an open neighborhood of $q_{\tau}^{(\bar{a}_{K})}$.
\end{condition}

To prove Theorem \ref{thm: consistency}, we first prove the following lemma:

\begin{lemma} \label{lemma: almost 0 root}
(i) For point treatment, the estimating equation for $q_{\tau}^{(a)}$ is 
\begin{align}
\label{eq:B lemma point trt almost 0 root}
    \Psi_N(q) = \frac{1}{N}\sum^{N}_{i = 1} \frac{\mathbbm{1}_{A_i = a}}{\hat{P}(A_i = a|L_i)} (\mathbbm{1}_{\tilde{Y}_i \leq q} - \tau) = 0. 
\end{align}

Under Assumptions \ref{asp: positivity} and \ref{asp: PS model correct} and Condition \ref{asp: compact L}, this estimating equation has an almost-zero root $\hat{q}^{(a)}_{\tau}$.

(ii) For time-varying treatment, the estimating equation for $q_{\tau}^{(\bar{a})}$ is
\begin{align}
\label{eq:B lemma time-vary trt almost 0 root}
\bar{\Psi}_N (q) = \frac{1}{N}\sum^{N}_{i = 1}\frac{\mathbbm{1}_{\bar{A}_{K, i} = \bar{a}_{K}}}{\prod^{K}_{k = 0} \hat{P} [A_{k, i} = a_k | \bar{A}_{k-1, i} = \bar{a}_{k - 1}, \bar{L}_{k, i}]} & \big( \mathbbm{1}_{\tilde{Y}_i \leq q} - \tau \big) = 0.  
\end{align}

Under Assumptions \ref{asp: positivity} and \ref{asp: PS model correct} and Condition \ref{asp: compact L}, this estimating equation has an almost-zero root $\hat{q}^{(\bar{a})}_{\tau}$.
\end{lemma}

\textit{Proof of Lemma \ref{lemma: almost 0 root}:}
\begin{align*}
\Psi_N(q) = & \frac{1}{N} \sum^{N}_{i = 1} \frac{\mathbbm{\mathbbm{1}}_{A_i = a}}{ \hat{P}(A_i = a |L_i)} (\mathbbm{1}_{\tilde{Y}_i \leq q} - \tau) \\
= & \frac{1}{N}  \sum^{N}_{i = 1} \frac{\mathbbm{1}_{A_i = a}}{ \hat{P}(A_i = a |L_i)} \mathbbm{1}_{\tilde{Y}_i \leq q}   - \frac{1}{N}  \sum^{N}_{i = 1} \frac{\mathbbm{1}_{A_i = a}}{ \hat{P}(A_i = a |L_i)} \tau.
\end{align*}
The second term does not depend on $q$. The first term does depend on $q$. 

$\Psi_N(q)$ is increasing in $q$. Small $q$ makes the first term smaller than the second term, so $\Psi_N(q)$  is negative. Large $q$ makes the first term larger than the second term, so $\Psi_N(q)$  is positive. Each time we increase $q$ to make $\Psi_N(q)$ larger by passing one of the $\tilde{Y}_i$ with $A_i = a$, $\Psi_N(q)$ will increase by $\mathbbm{1}_{A_i = a} /( N \cdot \hat{P}(A_i = a |L_i) )$. 

Assumption \ref{asp: PS model correct} assumes that the logistic regression model for $P(A_i = a |L_i)$ is correctly specified. By the asymptotic properties of the logistic regression estimator and Assumption~\ref{asp: compact L},
\begin{align*}
    \sup_{l} \|\hat{P}(A_i = a |L_i = l) - P(A_i = a | L_i = l)  \| \overset{P}{\to} 0,
\end{align*}
as $N \to \infty$.  

Let $\delta > 0$ be given. Choose $N$ such that for $\varepsilon > 0$ from Positivity Assumption \ref{asp: positivity}, for all $n \geq N$
\begin{align*}
    P \left( \sup_l \rvert \hat{P}(A_i = a|L_i = l) -P(A_i = a|L_i = l)| < \frac{\varepsilon}{2} \right ) > 1 - \delta.
\end{align*}
Since $P(A_i = a|L_i) > \varepsilon$ (Positivity Assumption \ref{asp: positivity}), the above equation leads to that for all $n \geq N$,
$$
P \left( \hat{P}(A_i = a|L_i) > \frac{\varepsilon}{2} \right ) > 1 - \delta.
$$
Therefore, for all $n \geq N$, the jumps of $\Psi_N(q)$, $\mathbbm{1}_{A_i = a} /( N \cdot \hat{P}(A_i = a |L_i) )$, are bounded by
\begin{align*}
    \frac{1}{N} \cdot \frac{1}{\varepsilon/2},
\end{align*}
with probability $> 1 - \delta$.

Consequently, equation \eqref{eq:B lemma point trt almost 0 root} has an almost zero root $\hat{q}^{(a)}_\tau$, because with probability $> 1 - \delta$, we can find $\hat{q}^{(a)}_\tau$ with
\begin{align*}
 \Big \rvert \frac{1}{N} \sum^{N}_{i = 1} \frac{\mathbbm{1}_{A_i = a}}{ \hat{P}(A_i = a |L_i)} (\mathbbm{1}_{\tilde{Y}_i \leq \hat{q}^{(a)}_\tau} - \tau) \Big \rvert \leq \frac{1}{N\cdot\varepsilon/2}  \to 0.
\end{align*}

Following a similar procedure, it can be shown that equation \eqref{eq:B lemma time-vary trt almost 0 root} has an almost zero root $\hat{q}^{(\bar{a})}_\tau$ for the time-varying setting.

\hfill $\square$

\textit{Proof of Theorem \ref{thm: consistency} using Lemma \ref{lemma: almost 0 root}:}

We will use Lemma 5.10 in Van der Vaart (2000) to prove Theorem \ref{thm: asymptotic normality with known PS}. From Lemma \ref{lemma: almost 0 root} in the previous subsection, $\Psi_N(q)$ has an almost zero root $\hat{q}^{(a)}_\tau$. Next, we show that for all $q$ fixed, $\Psi_N(q) \overset{P}{\to} F_{\tilde{Y}^{(a)}}(q) - \tau$.

\begin{align}
& \Psi_N(q) - (F_{\tilde{Y}^{(a)}}(q) - \tau ) \nonumber  \\
= &  \left( \frac{1}{N} \sum^{N}_{i =1} \frac{\mathbbm{1}_{A_i = a}}{\hat{P}(A_i = a|L_i)} \mathbbm{1}_{\tilde{Y}_i \leq q} - F_{\tilde{Y}^{(a)}}(q) \right)  - \left( \frac{1}{N} \sum^{N}_{i =1} \frac{\mathbbm{1}_{A_i = a}}{\hat{P}(A_i = a|L_i)} - 1 \right) \tau.  \label{eq: proof of thm 3}
\end{align}

Let $\delta > 0$ be given. For the first term in Equation \eqref{eq: proof of thm 3}, for $n \geq N$ from the proof of Lemma 1 and $\varepsilon$ from Positivity Assumption \ref{asp: positivity}, with probability $> 1 - \delta$,
\begin{align*}
& \Big \rvert \frac{1}{N} \sum^{N}_{i =1} \frac{\mathbbm{1}_{A_i = a}}{\hat{P}(A_i = a|L_i)} \mathbbm{1}_{\tilde{Y}_i \leq q} - F_{\tilde{Y}^{(a)}}(q) \Big \rvert \\
\leq & \Big \rvert \frac{1}{N} \sum^{N}_{i =1} \frac{\mathbbm{1}_{A_i = a}}{\hat{P}(A_i = a|L_i)}\mathbbm{1}_{\tilde{Y}_i \leq q}  - \frac{1}{N} \sum^{N}_{i =1} \frac{\mathbbm{1}_{A_i = a}}{P(A_i = a|L_i)}\mathbbm{1}_{\tilde{Y}_i \leq q} \Big \rvert  \\
 &  \qquad + \Big \rvert \frac{1}{N} \sum^{N}_{i =1} \frac{\mathbbm{1}_{A_i = a}}{P(A_i = a|L_i)} \mathbbm{1}_{\tilde{Y}_i \leq q}  - F_{\tilde{Y}^{(a)}}(q) \Big \rvert \\  
= & \Big \rvert \frac{1}{N} \sum^{N}_{i =1} \frac{P(A_i = a|L_i) - \hat{P}(A_i = a|L_i)}{\hat{P}(A_i = a|L_i)P(A_i = a|L_i)} \mathbbm{1}_{A_i = a} \mathbbm{1}_{\tilde{Y}_i \leq q}  \Big \rvert \\
& \qquad \qquad + \Big \rvert \frac{1}{N} \sum^{N}_{i =1} \frac{\mathbbm{1}_{A_i = a}}{P(A_i = a |L_i)}  \mathbbm{1}_{\tilde{Y}_i \leq q}  - F_{\tilde{Y}^{(a)}}(q) \Big \rvert  \\
\leq &  \frac{1}{N} \sum^{N}_{i =1}  \Big \rvert \frac{P(A_i = a|L_i)-\hat{P}(A_i = a |L_i)}{\hat{P}(A_i = a | L_i )P(A_i = a | L_i)}  \Big \rvert \\
& \qquad \qquad + \Big \rvert \frac{1}{N} \sum^{N}_{i =1} \frac{\mathbbm{1}_{A_i = a}}{P(A_i = a | L_i)}  \mathbbm{1}_{\tilde{Y}_i \leq q}  - F_{\tilde{Y}^{(a)}}(q) \Big \rvert \\
< & \frac{\delta/2}{\epsilon \cdot \epsilon/2} + \frac{\delta}{2} 
= \frac{\delta}{2} \left( \frac{2}{\epsilon^2} + 1\right).
\end{align*}
The first inequality uses the Triangular Inequality. For the first term in the fourth inequality, choose $N$ such that, for all $n \geq N$ with probability $>1 - \delta/2$, 
$\sup_l \rvert \hat{P}(A_i = a|L_i = l) -P(A_i = a|L_i = l)| < \varepsilon/2$ and $\sup_l \rvert \hat{P}(A_i = a|L_i = l) -P(A_i = a|L_i = l)| < \delta/2$. Then, $\hat{P}(A_i = a|L_i) \geq \varepsilon / 2$ with probability $> 1 - \delta/2$. For the second term in the fourth inequality, the Law of Large Numbers and Theorem \ref{thm: identification} imply that
\begin{align*}
    \frac{1}{N} \sum^{N}_{i =1} \frac{\mathbbm{1}_{A_i = a}}{P(A_i = a | L_i)}  \mathbbm{1}_{\tilde{Y}_i \leq q}  \overset{P}{\to} E \left( \frac{\mathbbm{1}_{A_i = a}}{P(A_i = a | L_i)}  \mathbbm{1}_{\tilde{Y}_i \leq q} \right)  =F_{\tilde{Y}^{(a)}}(q) ,
\end{align*}
so we can choose $N$ possibly even larger so that for all $n \geq N$, with probability $> 1 - \delta/2$,
\begin{align*}
    \Big \rvert \frac{1}{N} \sum^{N}_{i =1} \frac{\mathbbm{1}_{A_i = a}}{P(A_i = a | L_i)}  \mathbbm{1}_{\tilde{Y}_i \leq q}  - F_{\tilde{Y}^{(a)}}(q) \Big \rvert < \frac{\delta}{2}.
\end{align*}

In Equation \eqref{eq: proof of thm 3}, the second term can be bounded in a similar way:
\begin{align*}
& \Big \rvert \frac{1}{N} \sum^{N}_{i =1} \frac{\mathbbm{1}_{A_i = a}}{\hat{P}(A_i = a | L_i)}  - 1 \Big \rvert \\
\leq &  \Big \rvert \frac{1}{N} \sum^{N}_{i =1} \frac{\mathbbm{1}_{A_i = a}}{\hat{P}(A_i = a | L_i)}  - \frac{1}{N} \sum^{N}_{i =1} \frac{\mathbbm{1}_{A_i = a}}{P(A_i = a | L_i)} \Big \rvert  \\
& \qquad + \Big \rvert \frac{1}{N} \sum^{N}_{i =1} \frac{\mathbbm{1}_{A_i = a}}{P(A_i = a | L_i)}  - 1 \Big \rvert \\  
= & \Big \rvert \frac{1}{N} \sum^{N}_{i =1} \frac{P(A_i = a | L_i) - \hat{P}(A_i = a | L_i)}{\hat{P}(A_i = a | L_i)P(A_i = a | L_i)} \mathbbm{1}_{A_i = a} \Big \rvert + \Big \rvert \frac{1}{N} \sum^{N}_{i =1} \frac{\mathbbm{1}_{A_i = a}}{P(A_i = a | L_i)}  - 1 \Big \rvert  \\
< & \frac{\delta}{2} \left( \frac{2}{\epsilon^2} + 1\right). 
\end{align*}

Therefore, for all $n \geq N$, with probability $> 1 - \delta$,
\begin{align*}
  \rvert \Psi_N(q) - (F_{\tilde{Y}^{(a)}}(q) - \tau ) \rvert < \frac{\delta}{2} \left( \frac{2}{\epsilon^2} + 1\right) (1 + \tau).
\end{align*}
By chosing $\delta > 0$ small, this can be made arbitrarily small.
We conclude that, as $N \to \infty$, $\Psi_N(q) \overset{P}{\to} F_{\tilde{Y}^{(a)}}(q) - \tau$.

In conclusion, $q \to \Psi_N(q)$ is nondecreasing, $\hat{q}^{(a)}_{\tau}$ is an almost zero root of $\Psi_N(q)$ (because of Lemma \ref{lemma: almost 0 root}), $\Psi_N(q) \overset{P}{\to} F_{\tilde{Y}^{(a)}}(q) - \tau$, and $q^{(a)}_{\tau}$ is the root of $F_{\tilde{Y}^{(a)}}(q) - \tau$. Therefore, Lemma 5.10 in Van der Vaart (2000) implies that
\begin{align*}
     \hat{q}^{(a)}_{\tau}  \overset{P}{\to} q^{(a)}_{\tau}.
\end{align*}

Following a similar procedure, it can be shown that in the time-varying setting,
\begin{align*}
\hat{q}_{\tau}^{(\bar{a})} \overset{P}{\to} q_{\tau}^{(\bar{a})}. 
\end{align*} 
\hfill $\square$

\subsection{Proof of Theorem 3: asymptotic normality when the propensity score is known}

\textit{Proof:}

When the propensity score is known, the estimating equation for $\Tilde{q}^{(a)}_{\tau}$ is
\begin{align*}
    \Psi_N(q) = \frac{1}{N}\sum^N_{i=1} \frac{\mathbbm{1}_{A_i = a}}{P(A_i = a|L_i)}(\mathbbm{1}_{\tilde{Y}_i \leq q} - \tau).
\end{align*}

$\Psi_N(q)$ is a monotone increasing function in $q$. Similar to Lemma \ref{lemma: almost 0 root}, there exists an almost zero root  $\tilde{q}^{(a)}_{\tau}$ of $\Psi_N(q)$. We choose $\tilde{q}^{(a)}_{\tau}$ as the leftmost point where $\Psi_N(q)$ becomes $\geq 0$. Since jumps are bounded by $1/(N\varepsilon)$ with $N$ and $\varepsilon$ from Positivity Assumption \ref{asp: positivity}, it follows that $\left \rvert \Psi_N(\tilde{q}^{(a)}_{\tau}) \right \rvert \leq 1/(N \varepsilon)$. For this choice of $\tilde{q}^{(a)}_{\tau}$, $q$ is less than $\tilde{q}^{(a)}$, if and only if $\Psi_N(q) < 0$. Thus, 
\begin{align}
    P \left(\sqrt{N}(\tilde{q}^{(a)}_{\tau} - q^{(a)}_{\tau}) > \delta \right) & = P \left(\tilde{q}^{(a)}_{\tau} > q^{(a)}_{\tau} + \frac{\delta}{\sqrt{N}} \right) \hfill \nonumber \\
    & = P\left( \Psi_N \left(q^{(a)}_{\tau} + \frac{\delta}{\sqrt{N}} \right) < 0 \right) \nonumber \\
    & = P \left( \frac{1}{N}\sum^N_{i=1} \frac{\mathbbm{1}_{A_i = a}}{P(A_i = a|L_i)}\left(\mathbbm{1}_{\tilde{Y}_i \leq q^{(a)}_{\tau} + \frac{\delta}{\sqrt{N}}} - \tau \right) < 0 \right). \hfill \label{eq:B prob of gradient function}
\end{align}

To apply the Central Limit Theorem to \eqref{eq:B prob of gradient function}, we derive the mean and variance for the term \begin{align*}
    \frac{\mathbbm{1}_{A_i = a}}{P(A_i = a|L_i)}\left(\mathbbm{1}_{\tilde{Y}_i \leq q^{(a)}_{\tau} + \frac{\delta}{\sqrt{N}}} - \tau \right).
\end{align*}
As for the mean, using Theorem 1,
\begin{align*}
& E \left[  \frac{\mathbbm{1}_{A_i = a}}{P(A_i = a|L_i)}\left(\mathbbm{1}_{\tilde{Y}_i \leq q^{(a)}_{\tau} + \frac{\delta}{\sqrt{N}}} - \tau \right) \right]  =  F_{\tilde{Y}^{(a)}} \left( q^{(a)}_{\tau} + \frac{\delta}{\sqrt{N}} \right) - \tau .
\end{align*}
As for the variance,
\begin{align}
& \text{Var}\left(  \frac{\mathbbm{1}_{A_i = a}}{P(A_i = a|L_i)}\left(\mathbbm{1}_{\tilde{Y}_i \leq q^{(a)}_{\tau} + \frac{\delta}{\sqrt{N}}} - \tau \right) \right) \nonumber \\   
& = E \left[ \left[ \frac{\mathbbm{1}_{A_i = a}}{P(A_i = a|L_i)}\left(\mathbbm{1}_{\tilde{Y}_i \leq q^{(a)}_{\tau} + \frac{\delta}{\sqrt{N}}} - \tau \right)\right]^{2} \right] -  \left(F_{\tilde{Y}^{(a)}} \left( q^{(a)}_{\tau} + \frac{\delta}{\sqrt{N}} \right) - \tau \right)^2. \label{eq:B variance equation}
\end{align}
For the first term in equation \eqref{eq:B variance equation}, notice that
\begin{align*}
    & \left[ \frac{\mathbbm{1}_{A_i = a}}{P(A_i = a|L_i)}\left(\mathbbm{1}_{\tilde{Y}_i 
    \leq q^{(a)}_{\tau} + \frac{\delta}{\sqrt{N}}} - \tau \right)\right]^{2} <= \left( \frac{\mathbbm{1}_{A_i = a}}{P(A_i = a|L_i)} \right)^{2} , 
\end{align*} 
and because of Positivity Assumption \ref{asp: positivity}, $E\left[ \left( \frac{\mathbbm{1}_{A_i = a}}{P(A_i = a|L_i)} \right)^{2} \right] < \infty$. In addition, as $N \to \infty$,
\begin{align*}
    \left[ \frac{\mathbbm{1}_{A_i = a}}{P(A_i = a|L_i)}\left(\mathbbm{1}_{\tilde{Y}_i \leq q^{(a)}_{\tau} + \frac{\delta}{\sqrt{N}}} - \tau \right)\right]^{2} \to \left[ \frac{\mathbbm{1}_{A_i = a}}{P(A_i = a|L_i)}\left(\mathbbm{1}_{\tilde{Y}_i \leq q^{(a)}_{\tau}} - \tau \right)\right]^{2}.
\end{align*}
Therefore, Lebesgue's Dominated Convergence Theorem \citeplatex{athreya2006measure} implies that
\begin{align*}
 E \left[ \left[ \frac{\mathbbm{1}_{A_i = a}}{P(A_i = a|L_i)}\left(\mathbbm{1}_{\tilde{Y}_i \leq q^{(a)}_{\tau} + \frac{\delta}{\sqrt{N}}} - \tau \right)\right]^{2} \right] \to  E\left[ \left[ \frac{\mathbbm{1}_{A_i = a}}{P(A_i = a|L_i)}\left(\mathbbm{1}_{\tilde{Y}_i \leq q^{(a)}_{\tau}} - \tau \right)\right]^{2} \right].  
\end{align*}
For the second term in Equation \eqref{eq:B variance equation}, 
\begin{align*}
    F_{\tilde{Y}^{(a)}} \left( q^{(a)}_{\tau} + \frac{\delta}{\sqrt{N}} \right) - \tau \to F_{\tilde{Y}^{(a)}} \left( q^{(a)}_{\tau} \right) - \tau = 0, 
\end{align*}
since we assumed that $F_{\tilde{Y}^{(a)}} \big( q^{(a)}_{\tau} \big)$ is continuous in $q$ at $q_{\tau}^{(a)}$ (Assumption \ref{asp: continuous f(y) point trt}).
Combining with equation \eqref{eq:B variance equation}, we conclude that
\begin{align*}
& \text{Var}\left(  \frac{\mathbbm{1}_{A_i = a}}{P(A_i = a|L_i)}\left(\mathbbm{1}_{\tilde{Y}_i \leq q^{(a)}_{\tau} + \frac{\delta}{\sqrt{N}}} - \tau \right) \right)  
\to  E\left[ \left[ \frac{\mathbbm{1}_{A_i = a}}{P(A_i = a|L_i)}\left(\mathbbm{1}_{\tilde{Y}_i \leq q^{(a)}_{\tau}} - \tau \right)\right]^{2} \right] = \tilde{V}.
\end{align*}

Next, continuing from equation \eqref{eq:B prob of gradient function}, we subtract  $F_{\tilde{Y}^{(a)}} \left( q^{(a)}_{\tau} + {\delta}/{\sqrt{N}} \right) - \tau$ on both sides of the inequality to obtain

\begin{align}
& P \left( \sqrt{N}(\tilde{q}^{(a)}_{\tau} - q^{(a)}_{\tau}) > \delta \right)  \nonumber \\
& = P \bigg ( \frac{1}{N}\sum^N_{i=1} \frac{\mathbbm{1}_{A_i = a}}{P(A_i = a|L_i)} \big(\mathbbm{1}_{\tilde{Y}_i \leq q^{(a)}_{\tau} + \frac{\delta}{\sqrt{N}}} - \tau \big) < 0 \bigg ) \nonumber  \\
& = P \Bigg ( \frac{1}{N}\sum^N_{i=1} \frac{\mathbbm{1}_{A_i = a}}{P(A_i = a|L_i)}\Big(\mathbbm{1}_{\tilde{Y}_i \leq q^{(a)}_{\tau} + \frac{\delta}{\sqrt{N}}} - \tau \Big) - \Big(F_{\tilde{Y}^{(a)}} \Big( q^{(a)}_{\tau} + \frac{\delta}{\sqrt{N}} \Big) - \tau \Big)  \nonumber < \tau -F_{\tilde{Y}^{(a)}} \Big( q^{(a)}_{\tau} + \frac{\delta}{\sqrt{N}}  \Big) \Bigg) \nonumber \\
& = P \Bigg ( \sqrt{N} \Bigg( \frac{1}{N}\sum^N_{i=1} \frac{\mathbbm{1}_{A_i = a}}{P(A_i = a|L_i)}\Big(\mathbbm{1}_{\tilde{Y}_i \leq q^{(a)}_{\tau} + \frac{\delta}{\sqrt{N}}} - \tau \Big) - \bigg(F_{\tilde{Y}^{(a)}} \Big( q^{(a)}_{\tau} + \frac{\delta}{\sqrt{N}} \Big) - \tau \bigg) \Bigg) \nonumber \\ & \qquad \qquad < \sqrt{N} \bigg( \tau -F_{\tilde{Y}^{(a)}} \Big( q^{(a)}_{\tau} + \frac{\delta}{\sqrt{N}}  \Big) \bigg) \Bigg). \label{eq:B subtract for CLT}
\end{align}
For the term on the right hand side of  equation \eqref{eq:B subtract for CLT},
\begin{align*}
\sqrt{N}  \Big( \tau -F_{\tilde{Y}^{(a)}} \big( q^{(a)}_{\tau} + \frac{\delta}{\sqrt{N}} \big)  \Big) = & \sqrt{N} \Big( F_{\tilde{Y}^{(a)}} ( q^{(a)}_{\tau} ) - F_{\tilde{Y}^{(a)}} \big( q^{(a)}_{\tau} + \frac{\delta}{\sqrt{N}} \big)  \Big) \\
   = & \sqrt{N} \cdot f_{\tilde{Y}^{(a)}}(\dot{q}) \cdot \big(0 - \frac{\delta}{\sqrt{N}} \big)  \\
   \to & -\delta f_{\tilde{Y}^{(a)}}(q^{(a)}_{\tau}),
\end{align*}
where the second equality uses the Mean Value Theorem and $\dot{q}$ is a value between $q^{(a)}_{\tau}$ and $q^{(a)}_{\tau} + \delta / \sqrt{N}$, and convergence follows since we assumed that $f_{\tilde{Y}^{(a)}}(q)$ is continuous at $q_{\tau}^{(a)}$~(Assumption \ref{asp: continuous f(y) point trt}).

Applying the Triangular Central Limit Theorem \citep{athreya2006measure} to equation \eqref{eq:B subtract for CLT} results in 
\begin{align*}
    P \left( \sqrt{N}(\tilde{q}^{(a)}_{\tau} - q^{(a)}_{\tau}) > \delta \right) \to \Phi \left(-\frac{ \delta f_{\tilde{Y}^{(a)}} (q_{\tau}^{(a)}) }{\sqrt{\tilde{V}}} \right),
\end{align*}
where $\Phi(\cdot)$ is the Cumulative Distribution Function (CDF) of the standard normal distribution. It follows that
\begin{align*}
P( \sqrt{N}( \tilde{q}^{(a)}_{\tau} - q^{(a)}_{ \tau} ) \leq \delta )  \to \Phi\left(  \frac{ \delta f_{\tilde{Y}^{(a)}} (q^{(a)}_{ \tau} ) }{ \sqrt{ \tilde{V} } } \right).
\end{align*}
Hence, 
\begin{align*}
    \sqrt{N}( \tilde{q}^{(a)}_{\tau} - q^{(a)}_\tau) \rightarrow N\left(0, \frac{\tilde{V}}{ f^2_{\tilde{Y}^{(a)}}\left(q^{(a)}_{\tau} \right) } \right).
\end{align*}
\hfill $\square$

\subsection{Proof of Theorem 4: asymptotic normality when the propensity score is estimated}

\noindent 
\textit{Proof:} 

For asymptotic normality when the propensity score is estimated, we focus on the estimated $\tau$th quantile under treatment $a=1$. For $a=0$, the proof is similar. 

For $a=1$, the estimating equation for $\hat{q}^{(1)}_{\tau}$ (equation \eqref{eq: gradient function} in the main text) is 
\begin{align*}
    \Psi_N(q) = \frac{1}{N}\sum^N_{i=1} \frac{\mathbbm{1}_{A_i = 1}}{p_{\hat{\theta}}(A_i = 1|L_i)}(\mathbbm{1}_{\tilde{Y}_i \leq q} - \tau).
\end{align*}
$\Psi_N(q)$ is a monotone increasing function in $q$. From Lemma \ref{lemma: almost 0 root}, there exists an almost zero root  $\hat{q}^{(1)}_{\tau}$ of $\Psi_N(q)$. We choose $\hat{q}^{(1)}_{\tau}$ as the left-most point where $\Psi_N(q)$ becomes $\geq 0$. For this choise of $\hat{q}^{(1)}_{\tau}$,  $\hat{q}^{(1)}_\tau$ is greater than $q$, if and only if $\Psi_N(q) < 0$. Thus, 
\begin{eqnarray}
\lefteqn{P \left(\sqrt{N}(\hat{q}^{(1)}_{\tau} - q^{(1)}_{\tau}) > \delta \right)}\nonumber\\
 & = & P \left(\hat{q}^{(1)}_{\tau} > q^{(1)}_{\tau} + \frac{\delta}{\sqrt{N}} \right) \nonumber\\
    & = &P\left( \Psi_N \left(q^{(1)}_{\tau} + \frac{\delta}{\sqrt{N}} \right) < 0 \right) \nonumber\\
    & = &P \left( \frac{1}{N}\sum^N_{i=1} \frac{\mathbbm{1}_{A_i = 1}}{p_{\hat{\theta}}(A_i = 1|L_i)}\left(\mathbbm{1}_{\tilde{Y}_i \leq q^{(1)}_{\tau} + \frac{\delta}{\sqrt{N}}} - \tau \right) < 0 \right).\label{probability}
\end{eqnarray}

A Taylor expansion leads to
\begin{align}
\label{pf thm4: MVT}
& \frac{1}{N}\sum^N_{i=1} \frac{\mathbbm{1}_{A_i = 1}}{p_{\hat{\theta}}(A_i = 1|L_i)}\left(\mathbbm{1}_{\tilde{Y}_i \leq q^{(1)}_{\tau} + \frac{\delta}{\sqrt{N}}} - \tau \right) \nonumber\\
&=\frac{1}{N}\sum^N_{i=1} \frac{\mathbbm{1}_{A_i = 1}}{p_{\theta^*}(A_i = 1|L_i)}\left(\mathbbm{1}_{\tilde{Y}_i \leq q^{(1)}_{\tau} + \frac{\delta}{\sqrt{N}}} - \tau \right)  \nonumber \\
& \; \; +\frac{1}{N}\sum^N_{i=1}  \left.\frac{\partial}{\partial\theta}\right|_{\dot{\theta}}\frac{\mathbbm{1}_{A_i = 1}}{p_{\theta}(A_i = 1|L_i)}\left(\mathbbm{1}_{\tilde{Y}_i \leq q^{(1)}_{\tau} + \frac{\delta}{\sqrt{N}}} - \tau \right)(\hat{\theta}-\theta^*),
\end{align}
for some $\dot{\theta}$ between $\hat{\theta}$ and $\theta^*$.
Next, notice that $\hat{\theta}$ is estimated by maximum partial likelihood, and it solves partial score equations of the form
\begin{equation*}
\mathbbm{P}_NU_2(A,L;\theta)=0,
\end{equation*}
where $\mathbbm{P}_N$ is the empirical distribution, $\mathbbm{P}_{N} f (A, L)= N^{-1} \sum ^{N}_{i =1} f(A_i, L_i)$ for observations  $(A_1, L_1), ..., (A_N, L_N)$, and $U_2(A,L;\theta)$ is the partial score function for $\theta$.

Then, from theory on unbiased estimating equations, Theorem~5.21 in \citetlatex{van2000asymptotic} implies that
\begin{equation}
\label{pf thm4: thm 5.21}
\sqrt{N}(\hat{\theta}-\theta^*) = I\bigl(\theta^*\bigr)^{-1}\sqrt{N}\frac{1}{N}\sum_{i=1}^N U_2(\theta^*)+o_P(1),
\end{equation}
where $I(\theta^*) = -E\left(\left.\frac{\partial}{\partial \theta}\right|_{\theta^*} U_2(\theta)\right)$ is the partial Fisher information for $\theta$ from partial likelihood theory on estimation of $\theta$, since $U_2$ is the partial score for $\theta$. Combining equations \eqref{pf thm4: MVT} and \eqref{pf thm4: thm 5.21}, it follows that
\begin{align}
 & \lefteqn{\sqrt{N}\frac{1}{N}\sum^N_{i=1} \frac{\mathbbm{1}_{A_i = 1}}{p_{\hat{\theta}}(A_i = 1|L_i)}\left(\mathbbm{1}_{\tilde{Y}_i \leq q^{(1)}_{\tau} + \frac{\delta}{\sqrt{N}}} - \tau \right)} \nonumber\\
& = \sqrt{N}\frac{1}{N}\sum^N_{i=1} \frac{\mathbbm{1}_{A_i = 1}}{p_{\theta^*}(A_i = 1|L_i)}\left(\mathbbm{1}_{\tilde{Y}_i \leq q^{(1)}_{\tau} + \frac{\delta}{\sqrt{N}}} - \tau \right)  \nonumber \\
& + \left( \frac{1}{N}\sum^N_{i=1}  \left.\frac{\partial}{\partial\theta}\right|_{\dot{\theta}}\frac{\mathbbm{1}_{A_i = 1}}{p_{\theta}(A_i = 1|L_i)}\left(\mathbbm{1}_{\tilde{Y}_i \leq q^{(1)}_{\tau} + \frac{\delta}{\sqrt{N}}} - \tau \right) \right) \cdot \left(I\bigl(\theta^*\bigr)^{-1}\sqrt{N}\frac{1}{N}\sum_{i=1}^n U_2(\theta^*)+o_P(1)\right). \label{eq: combining Van der vaart}
\end{align}

Since the propensity score is modeled with logistic regression,
\begin{equation*}
p_{\theta}(A_i = 1|L_i)=\frac{e^{\theta^{\top}L_i}}{1 + e^{\theta^\top L_i}}.
\end{equation*}
Hence 
\begin{align}
\label{pf thm4: PS derivation}
\frac{\partial}{\partial\theta}\frac{1}{p_{\theta}(A_i = 1|L_i)} &  = \frac{\partial}{\partial\theta}  \left(  \frac{1}{e^{\theta^\top L_i}} + 1 \right) \nonumber \\
& =- L_i^\top  \frac{1}{e^{\theta^\top L_i}}  \nonumber   \\
& =- L_i^\top \left(\frac{1 + e^{\theta^\top L_i}}{e^{\theta^\top L_i}} - 1 \right)  \nonumber  \\
& = L_i^\top  \left(1 - \frac{1}{p_{\theta}(A_i = 1|L_i)}  \right) \nonumber \\
& = L_i^\top   \frac{p_{\theta}(A_i = 1|L_i) - 1 }{p_{\theta}(A_i = 1|L_i)}.  
\end{align}

Therefore, in equation \eqref{eq: combining Van der vaart}

\begin{eqnarray}
\lefteqn{\frac{1}{N}\sum^N_{i=1}  \left.\frac{\partial}{\partial\theta}\right|_{\dot{\theta}}\frac{\mathbbm{1}_{A_i = 1}}{p_{\theta}(A_i = 1|L_i)}\left(\mathbbm{1}_{\tilde{Y}_i \leq q^{(1)}_{\tau} + \frac{\delta}{\sqrt{N}}} - \tau \right)}\nonumber\\
&=& \frac{1}{N} \sum_{i=1}^{N} L_i^\top \left(p_{\dot{\theta}}(A_i = 1|L_i) - 1\right) \frac{\mathbbm{1}_{A_i = 1}}{p_{\dot{\theta}}(A_i = 1|L_i)} \left( \mathbbm{1}_{\tilde{Y}_i \leq q_{\tau}^{(1)} + \frac{\delta}{\sqrt{N}}} - \tau \right) \nonumber\\
&=& \frac{1}{N} \sum_{i=1}^{N} L_i^\top \left(p_{\theta^*}(A_i = 1|L_i) - 1\right) \frac{\mathbbm{1}_{A_i = 1}}{p_{\theta^*}(A_i = 1|L_i)} \left( \mathbbm{1}_{\tilde{Y}_i \leq q_{\tau}^{(1)} + \frac{\delta}{\sqrt{N}}} - \tau \right) + o_{P}(1)\nonumber\\
&\overset{P}{\to} & E\left(L_i^\top \left(p_{\theta^*}(A_i = 1|L_i) - 1\right) \frac{\mathbbm{1}_{A_i = 1}}{p_{\theta^*}(A_i = 1|L_i)} \left( \mathbbm{1}_{\tilde{Y}_i \leq q_{\tau}^{(1)} } - \tau \right) \right). \label{Ddef}
\end{eqnarray}
The second equality in Equation \eqref{Ddef} follows from a Taylor expansion with $\ddot{\theta}$ between $\dot{\theta}$ and $\theta^*$:
\begin{align}
\label{pf thm4: Taylor expan}
& \frac{1}{N} \sum_{i=1}^{N} L_i^\top \left(p_{\dot{\theta}}(A_i = 1|L_i) - 1\right) \frac{\mathbbm{1}_{A_i = 1}}{p_{\dot{\theta}}(A_i = 1|L_i)} \left( \mathbbm{1}_{\tilde{Y}_i \leq q_{\tau}^{(1)} + \frac{\delta}{\sqrt{N}}} - \tau \right) \nonumber \\
& = \frac{1}{N} \sum_{i=1}^{N} L_i^\top \left(p_{\theta^*}(A_i = 1|L_i) - 1\right) \frac{\mathbbm{1}_{A_i = 1}}{p_{\theta^*}(A_i = 1|L_i)} \left( \mathbbm{1}_{\tilde{Y}_i \leq q_{\tau}^{(1)} + \frac{\delta}{\sqrt{N}}} - \tau \right)  \nonumber \\
& + \frac{1}{N} \sum_{i=1}^{N}  \underbrace{\frac{\partial}{\partial \theta} \bigg|_{\ddot{\theta}}\left(L_i^\top \left(p_{\theta}(A_i = 1|L_i) - 1\right)  \frac{\mathbbm{1}_{A_i = 1}}{p_{\theta}(A_i = 1|L_i)} \right)}_\text{factor 1} \underbrace{\left( \mathbbm{1}_{\tilde{Y}_i \leq q_{\tau}^{(1)} + \frac{\delta}{\sqrt{N}}} - \tau \right)}_\text{factor 2} \underbrace{ \left( \dot{\theta} - \theta^* \right)}_\text{factor 3}.
\end{align}
For factor 1, using equation \eqref{pf thm4: PS derivation} leads to
\begin{align*}
 \frac{\partial}{\partial \theta} \bigg|_{\ddot{\theta}} \left( L_i^\top \left(p_{\theta}(A_i = 1|L_i) - 1\right) \frac{\mathbbm{1}_{A_i = 1}}{p_{\theta}(A_i = 1|L_i)} \right) 
& = \frac{\partial}{\partial \theta} \bigg|_{\ddot{\theta}} \left( L_i^\top  \mathbbm{1}_{A_i = 1} \left( 1 - \frac{1}{p_{\theta}(A_i = 1|L_i)} \right) \right)\\
& = L_i L_i^\top \mathbbm{1}_{A_i = 1} \left( \frac{1}{p_{\ddot{\theta}}(A_i = 1|L_i)} - 1\right).
\end{align*}
A procedure similar to the proof of Lemma 1 leads to that choosing $\epsilon > 0$ from Positivity Assumption \ref{asp: positivity}, there exist a $\delta > 0$ that for all $n \geq N$, the term $1/{p_{\ddot{\theta}}(A_i = 1|L_i)}$ is bounded by $2/\epsilon$ with probability $> 1 - \delta$. Since
\begin{align*}
\left| L_i L_i^\top \mathbbm{1}_{A_i = 1} \left( \frac{1}{p_{\ddot{\theta}}(A_i = 1|L_i)} - 1 \right) \right| = \left| L_i L_i^\top \mathbbm{1}_{A_i = 1} \left( \frac{1 - p_{\ddot{\theta}}(A_i = 1|L_i)}{p_{\ddot{\theta}}(A_i = 1|L_i)} \right) \right| < \frac{ \abs{L_i L_i^\top} }{p_{\ddot{\theta}}(A_i = 1|L_i)},
\end{align*}
let $C = 2\abs{L_i L_i^\top}/\epsilon$, then factor 1 from Equation \eqref{pf thm4: Taylor expan} is bounded by $C$ with probability $> 1 - \delta$ because $L_i$ belongs to a compact set (Assumption \ref{asp: compact L}).

Factor 2 from Equation \eqref{pf thm4: Taylor expan}, $\mathbbm{1}_{\tilde{Y}_i \leq q_{\tau}^{(1)} + \delta/\sqrt{N}} - \tau$, is bounded by 1. For factor 3, $\dot{\theta}$ converges in probability to $\theta^*$. Therefore, the second term of the expression in equation \eqref{pf thm4: Taylor expan} converges in probability to 0. The second equality in equation \eqref{Ddef} follows.

Write $D^\top$ for the right hand side of equation \eqref{Ddef}, that is, 
\begin{align*}
    D^\top = E\left(  L_i^\top \left(p_{\theta^*}(A_i = 1|L_i) - 1\right) \frac{\mathbbm{1}_{A_i = 1}}{p_{\theta^*}(A_i = 1|L_i)} \left( \mathbbm{1}_{\tilde{Y}_i \leq q_{\tau}^{(1)} } - \tau \right) \right).
\end{align*}
Combining equation \eqref{eq: combining Van der vaart} and \eqref{Ddef} leads to
\begin{eqnarray*}
 \lefteqn{\sqrt{N}\frac{1}{N}\sum^N_{i=1} \frac{\mathbbm{1}_{A_i = 1}}{p_{\hat{\theta}}(A_i = 1|L_i)}\left(\mathbbm{1}_{\tilde{Y}_i \leq q^{(1)}_{\tau} + \frac{\delta}{\sqrt{N}}} - \tau \right)}\nonumber\\
&=&\sqrt{N}\frac{1}{N}\sum^N_{i=1} \frac{\mathbbm{1}_{A_i = 1}}{p_{\theta^*}(A_i = 1|L_i)}\left(\mathbbm{1}_{\tilde{Y}_i \leq q^{(1)}_{\tau} + \frac{\delta}{\sqrt{N}}} - \tau \right)\nonumber\\
&&+\left(D^\top + o_P(1)\right)\cdot \left(I\bigl(\theta^*\bigr)^{-1}\sqrt{N}\frac{1}{N}\sum_{i=1}^n U_2(\theta^*)+o_P(1)\right)\nonumber\\
&=&\sqrt{N}\frac{1}{N}\sum^N_{i=1} \frac{\mathbbm{1}_{A_i = 1}}{p_{\theta^*}(A_i = 1|L_i)}\left(\mathbbm{1}_{\tilde{Y}_i \leq q^{(1)}_{\tau} + \frac{\delta}{\sqrt{N}}} - \tau \right)\nonumber\\
&&+D^\top I\bigl(\theta^*\bigr)^{-1}\sqrt{N}\frac{1}{N}\sum_{i=1}^n U_2(\theta^*)+o_P(1).\end{eqnarray*}
Combining with equation~\eqref{probability}, we conclude that
\begin{eqnarray}
\lefteqn{P\left(\sqrt{N}(\hat{q}^{(1)}_{\tau} - q^{(1)}_{\tau}) > \delta \right)}\nonumber\\
& = & P\Biggl(\sqrt{N}\frac{1}{N}\sum^N_{i=1} \frac{\mathbbm{1}_{A_i = 1}}{p_{\theta^*}(A_i = 1|L_i)}\left(\mathbbm{1}_{\tilde{Y}_i \leq q^{(1)}_{\tau} + \frac{\delta}{\sqrt{N}}} - \tau \right) \nonumber\\
&& \qquad +D^\top I\bigl(\theta^*\bigr)^{-1}\sqrt{N}\frac{1}{N}\sum_{i=1}^n U_2(\theta^*)+o_P(1) < 0 \Biggr).\label{CLTstarter}
\end{eqnarray}

To apply the Triangular Central Limit Theorem \citep{athreya2006measure}, we derive the mean, variance, and covariance for the left hand side of the inequality inside the $P$ in equation~\eqref{CLTstarter}. The variance of the whole terms is 
\begin{align*}
&    \text{Var}\left(\frac{\mathbbm{1}_{A_i = 1}}{p_{\theta^*}(A_i = 1|L_i)}\left(\mathbbm{1}_{\tilde{Y}_i \leq q^{(1)}_{\tau} + \frac{\delta}{\sqrt{N}}} - \tau \right)\right)  \\
& + 2D^\top I\bigl(\theta^*\bigr)^{-1} {\rm COV}\left(  \frac{\mathbbm{1}_{A_i = 1}}{p_{\theta^*}(A_i = 1|L_i)}\left(\mathbbm{1}_{\tilde{Y}_i \leq q^{(1)}_{\tau} + \frac{\delta}{\sqrt{N}}} - \tau \right),U_2 \right) \\
& + \text{Var}\left( D^\top I\bigl(\theta^*\bigr)^{-1} U_2(\theta^*) \right).
\end{align*}

For the term 
\begin{equation*}
    \frac{\mathbbm{1}_{A_i = 1}}{p_{\theta^*}(A_i = 1|L_i)}\left(\mathbbm{1}_{\tilde{Y}_i \leq q^{(1)}_{\tau} + \frac{\delta}{\sqrt{N}}} - \tau \right),
\end{equation*}
the mean and variance are already derived in the proof of Theorem \ref{thm: asymptotic normality with known PS}. The mean is shown to converge to $\delta f_{\tilde{Y}^{(1)}}(q^{(1)}_{\tau})$. The variance is shown to converge to 
\begin{align*}
        \tilde{V} = E\left[ \left[ \frac{\mathbbm{1}_{A_i = 1}}{p_{\theta^*}(A_i = 1|L_i)}\left(\mathbbm{1}_{\tilde{Y}_i \leq q^{(1)}_{\tau}} - \tau \right)\right]^{2} \right].
\end{align*}

For the term 
\begin{align*}
D^\top I\bigl(\theta^*\bigr)^{-1} U_2(\theta^*),
\end{align*}
mean is 0 and variance is $D^\top I\bigl(\theta^*\bigr)^{-1}D$.
\

For the covariance, since $E(U_2)  = 0$,
\begin{eqnarray}
\lefteqn{{\rm COV}\left(  \frac{\mathbbm{1}_{A_i = 1}}{p_{\theta^*}(A_i = 1|L_i)}\left(\mathbbm{1}_{\tilde{Y}_i \leq q^{(1)}_{\tau} + \frac{\delta}{\sqrt{N}}} - \tau \right),U_2 \right)}\nonumber\\
&=&  E \left(\frac{\mathbbm{1}_{A_i = 1}}{p_{\theta^*}(A_i = 1|L_i)}\left(\mathbbm{1}_{\tilde{Y}_i \leq q^{(1)}_{\tau} + \frac{\delta}{\sqrt{N}}} - \tau \right)\cdot U_2\right). \label{cov}  
\end{eqnarray}
Similar to the reasoning for the variance in the proof of Theorem \ref{thm: asymptotic normality with known PS}, the integrand is bounded by an integrable function. 
In addition, as $N \to \infty$,
\begin{equation*}
    \frac{\mathbbm{1}_{A_i = 1}}{p_{\theta^*}(A_i = 1|L_i)}\left(\mathbbm{1}_{\tilde{Y}_i \leq q^{(1)}_{\tau} + \frac{\delta}{\sqrt{N}}} - \tau \right) \cdot U_2 \stackrel{\text{a.s.}}{\to}  \frac{\mathbbm{1}_{A_i = 1}}{p_{\theta^*}(A_i = 1|L_i)}\left(\mathbbm{1}_{\tilde{Y}_i \leq q^{(1)}_{\tau}} - \tau \right)\cdot U_2.
\end{equation*}
Therefore, Lebesgue's Dominated Convergence Theorem \citep{athreya2006measure} implies that
\begin{eqnarray*}
\lefteqn{{\rm COV}\left(  \frac{\mathbbm{1}_{A_i = 1}}{p_{\theta^*}(A_i = 1|L_i)}\left(\mathbbm{1}_{\tilde{Y}_i \leq q^{(1)}_{\tau} + \frac{\delta}{\sqrt{N}}} - \tau \right),U_2 \right)} \\
& = & E \left(\frac{\mathbbm{1}_{A_i = 1}}{p_{\theta^*}(A_i = 1|L_i)}\left(\mathbbm{1}_{\tilde{Y}_i \leq q^{(1)}_{\tau} + \frac{\delta}{\sqrt{N}}} - \tau \right)\cdot U_2\right)\\
& \to&  E\left(\frac{\mathbbm{1}_{A_i = 1}}{p_{\theta^*}(A_i = 1|L_i)}\left(\mathbbm{1}_{\tilde{Y}_i \leq q^{(1)}_{\tau}} - \tau \right)\cdot U_2 \right)=-D,  
\end{eqnarray*}
since $U_2 \mathbbm{1}_{A_i = 1}=L_i \left(A_i - p_{\theta^*}(A_i = 1|L_i)\right)\mathbbm{1}_{A_i = 1} =L_i \left(1 - p_{\theta^*}(A_i = 1|L_i)\right)\mathbbm{1}_{A_i = 1}$. Therefore, the variance of the left hand side of the inequality inside $P$ in equation~\eqref{CLTstarter} converges to
\begin{eqnarray}
V
&=&\tilde{V}-2D^\top I\bigl(\theta^*\bigr)^{-1}D + D^\top I\bigl(\theta^*\bigr)^{-1}D \nonumber\\
&=&\tilde{V}-D^\top I\bigl(\theta^*\bigr)^{-1}D.
\end{eqnarray}

Now, to apply the Triangular Central Limit Theorem to equation~\eqref{CLTstarter}, we subtract  $\delta f_{\tilde{Y}^{(1)}}(q^{(1)}_{\tau})$ on both sides of the inequality to obtain
\begin{eqnarray*}
\lefteqn{P \left( \sqrt{N}(\hat{q}^{(1)}_{\tau} - q^{(1)}_{\tau}) > \delta \right)}\\
& = &P\Biggl(\sqrt{N}\frac{1}{N}\sum^N_{i=1} \frac{\mathbbm{1}_{A_i = 1}}{p_{\theta^*}(A_i = 1|L_i)}\left(\mathbbm{1}_{\tilde{Y}_i \leq q^{(1)}_{\tau} + \frac{\delta}{\sqrt{N}}} - \tau \right)-\delta f_{\tilde{Y}^{(1)}}(q^{(1)}_{\tau}) \nonumber\\
&&+D^\top I\bigl(\theta^*\bigr)^{-1}\sqrt{N}\frac{1}{N}\sum_{i=1}^n U_2(\theta^*)+o_P(1) < -\delta f_{\tilde{Y}^{(1)}}(q^{(1)}_{\tau}) \Biggr).
\end{eqnarray*}
Applying the Triangular Central Limit Theorem leads to
\begin{align*}
    P \left( \sqrt{N}(\hat{q}^{(1)}_{\tau} - q^{(1)}_{\tau}) > \delta \right) \to \Phi \left(-\frac{ \delta f_{\tilde{Y}^{(1)}} (q_{\tau}^{(1)}) }{\sqrt{V}} \right),
\end{align*}
where $\Phi(\cdot)$ is the CDF of the standard normal distribution. This implies that
\begin{equation*}
P( \sqrt{N}( \hat{q}^{(1)}_{\tau} - q^{(1)}_{ \tau} ) \leq \delta )  \to \Phi\left(  \frac{ \delta f_{\tilde{Y}^{(1)}} (q^{(1)}_{ \tau} ) }{ \sqrt{V} } \right).
\end{equation*}

It follows that
\begin{equation*}
    \sqrt{N}( \hat{q}^{(1)}_{\tau} - q^{(1)}_\tau) \rightarrow N\left(0, \frac{V}{ f^2_{\tilde{Y}^{(1)}}\left(q^{(1)}_{\tau} \right) } \right).
\end{equation*}

Notice that since $D^\top I\bigl(\theta^*\bigr)^{-1}D$ is positive semi-definite, estimating the nuisance parameter $\theta$ leads to a variance of $\hat{q}^{(1)}_{\tau}$ that is at most the variance $V$ one would obtain by using the known $\theta^*$ (Theorem 3). This is also seen for IPTW to estimate the mean \citeplatex{robins1994estimation}.\\

\hfill $\square$

\section{Additional simulation results}

\subsection{Simulation scenario of a time-varying setting}

\begin{figure}[H]
  \centerline{\includegraphics[width=5 in]{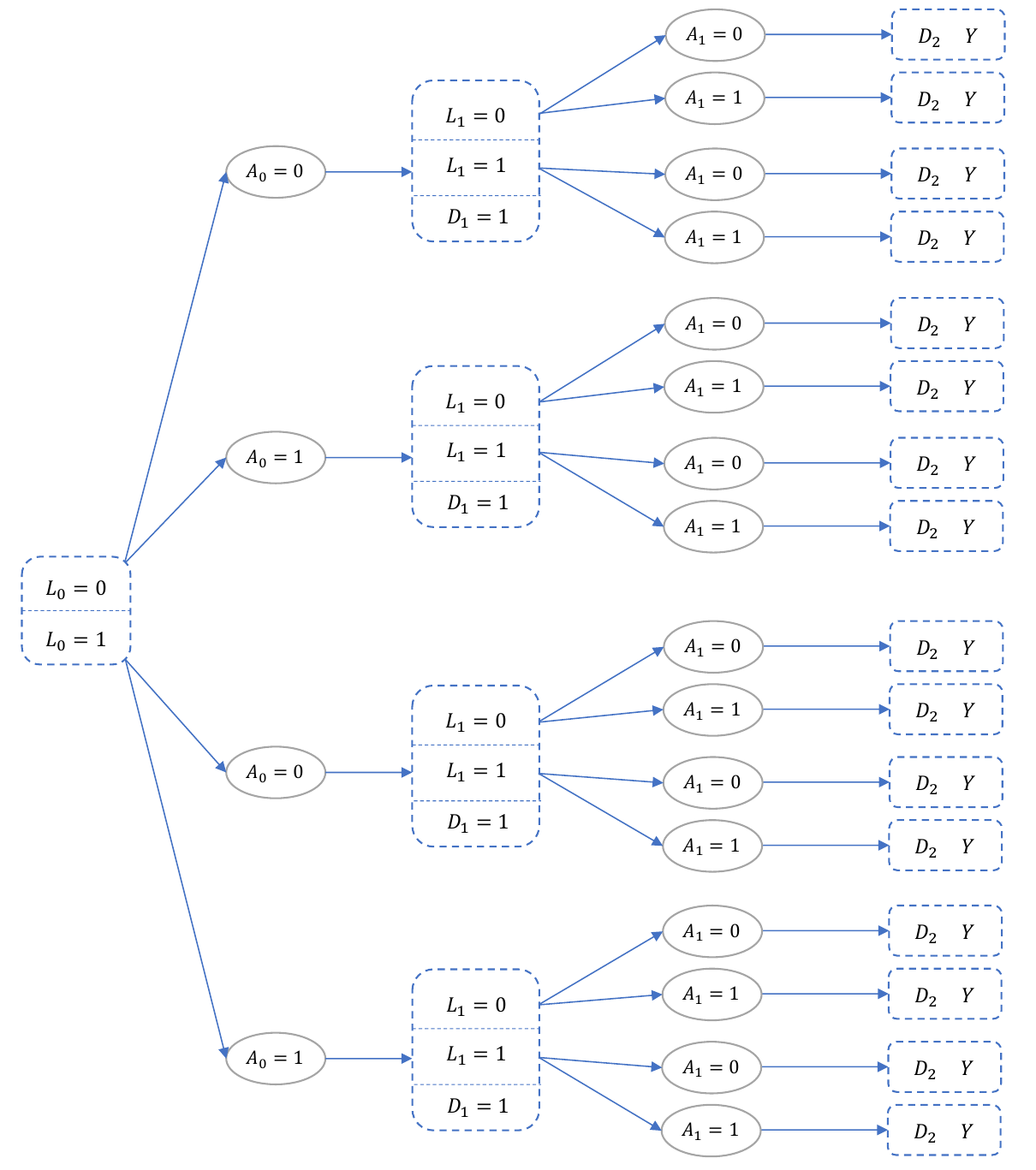}}
  \caption{Simulation scenario of a time-varying setting with observational data}
\label{f:timevar}
\end{figure}

\subsection{True survival-incorporated median in the point treatment simulation setting} 
\label{apdx: derivation point trt sim}

For the true survival-incorporated median in the point treatment setting, the distribution of the counterfactual ranked composite outcome can be derived mathematically. For treatment $a = 0$,
 \begin{align*}
F_{\tilde{Y}^{(0)}}(y)   = & P(L_i = 0) \cdot P(D^{(0)}_i = 1|L_i = 0) + P(L_i = 1) \cdot P(D^{(0)}_i = 1|L_i = 1)\\
& +  P(L_i = 0) \cdot P(D^{(0)}_i = 0|L_i = 0) \cdot F_{\tilde{Y}^{(0)} | D^{(0)} = 0, L = 0} (y)  \\
& + P(L_i = 1)  \cdot P(D^{(0)}_i = 0|L_i = 1) \cdot  F_{\tilde{Y}^{(0)} | D^{(0)} = 0, L = 1} (y)  \\
= & 0.4 \cdot 0.10  + 0.6 \cdot 0.16  + 0.4 \cdot 0.90 F_{\tilde{Y}^{(0)} | D^{(0)} = 0, L = 0}(y) + 0.6 \cdot 0.84 F_{\tilde{Y}^{(0)} | D^{(0)} = 0, L = 1}(y) \\
= & 0.136 + 0.36 F_{\tilde{Y}^{(0)} | D^{(0)} = 0, L = 0}(y) + 0.504 F_{\tilde{Y}^{(0)} | D^{(0)} = 0, L = 1}(y),
\end{align*}
where $F_{\tilde{Y}^{(0)} | D^{(0)} = 0, L = 0}(y) \sim N(0, 1)$ and  $F_{\tilde{Y}^{(0)} | D^{(0)} = 0, L = 1}(y) \sim N(3, 1)$.

By the definition of quantiles, solving the equation $F_{\tilde{Y}^{(0)}}(y) = 0.5$ for $y$ leads to the true survival-incorporated median $q^{(0)}_{0.5}$.

Similarly, we derive the distribution of the composite outcome under treatment $a = 1$:
\begin{align*}
& F_{\tilde{Y}^{(1)}}(y) = 0.068 + 0.38 F_{\tilde{Y}^{(1)} | D^{(1)} = 0, L = 0}(y) + 0.552 F_{\tilde{Y}^{(1)} | D^{(1)} = 0, L = 1}(y), 
\end{align*}
and solving the equation $F_{\tilde{Y}^{(1)}}(y) = 0.5$ for $y$ leads to the true survival-incorporated median $q^{(1)}_{0.5}$

Derivation of the median in the survivors is similar. The difference is that now the CDF of interest is conditional on survival $(D=0)$. For example, for $a=0$, the probability of survival is $1 - 0.136 = 0.864$, so the CDT should be calculated conditional on this probability:

\begin{align*}
F_{\tilde{Y}^{(0)}|D^{(0)} = 0}(y) & = \frac{360}{864} F_{\tilde{Y}^{(0)} | D^{(0)} = 0, L = 0}(y) + \frac{504}{864} F_{\tilde{Y}^{(0)} | D^{(0)} = 0, L = 1}(y) \\
F_{\tilde{Y}^{(1)}|D^{(1)} = 0}(y) & =  \frac{380}{932} F_{\tilde{Y}^{(1)} | D^{(1)} = 0, L = 0}(y) + \frac{552}{932} F_{\tilde{Y}^{(1)} | D^{(1)} = 0, L = 1}(y).
\end{align*}

\subsection{True survival-incorporated median in the time-varying simulation setting}
\label{apdx: derivation tv trt sim}

Similar to the point treatment setting, the true survival-incorporated median $q^{(0,0)}_{0.5}$ can be derived as follows.  The CDF of the potential composite outcome $\tilde{Y}^{(0, 0)}$ under treatment regimen $(0, 0)$ is
\begin{align*}
& F_{\tilde{Y}^{(0, 0)}}(y)  =  \sum_{l_0 \in (0, 1)} P(L_{0, i} = l_0) P(D_{1, i}^{(0)} = 1|L_{0, i} = l_0) \\
+ & \sum_{l_0 \in (0, 1)} P(L_{0, i} = l_0) P(D_{1, i}^{(0)} = 0|L_{0, i} = l_0) \\
\cdot &  \sum_{l_1 \in (0, 1)} P(L_{1, i}^{(0)} = l_1| D_{1, i}^{(0)} = 0, L_{0, i}=l_0) P(D_{2, i}^{(0, 0)} = 1| L_{1, i}^{(0)} = l_1, D_{1, i}^{(0)} = 0, L_{0, i} = l_0) \\
+ & \sum_{l_0 \in (0, 1)} P(L_{0, i} = l_0) P(D_{1, i}^{(0)} = 0|L_{0, i} = l_0) \\
\cdot & \sum_{l_1 \in (0, 1)} P(L_{1, i}^{(0)} = l_1 | D_{1, i}^{(0)} = 0, L_{0, i} = l_0) P(D_{2, i}^{(0, 0)} = 0|L_{1, i}^{(0)} = l_1, D_{1, i}^{(0)} = 0, L_{0, i} = l_0)  \\
& \qquad \cdot   F_{\tilde{Y}^{(0, 0)} | D_{2}^{(0, 0)} = 0, L_1^{(0)} = l_1, L_0 = l_0} (y)  .
\end{align*}
Plugging in the conditional probabilities from the simulated setting leads to:
\begin{align*}
F_{\tilde{Y}^{(0, 0)}}(y) = & 0.170 \\
& + 0.257 F_{\tilde{Y}^{(0,0)} | D_{2}^{(0, 0)} = 0, L_0 = 0, L_1^{(0)} = 0}(y)  + 0.092 F_{\tilde{Y}^{(0,0)} | D_{2}^{(0, 0)} = 0, L_0 = 0, L_1^{(0)} = 1}(y) \\
& + 0.133 F_{\tilde{Y}^{(0,0)} | D_{2}^{(0, 0)} = 0, L_0 = 1, L_1^{(0)} = 0}(y) + 0.348 F_{\tilde{Y}^{(0,0)} | D_{2}^{(0, 0)} = 0, L_0 = 1, L_1^{(0)} = 1}(y).
\end{align*}
Solving the equation $F_{\tilde{Y}^{(0, 0)}}(y) = 0.5$ leads to the true survival-incorporated median $q^{(0,0)}_{0.5}$. The derivations for other treatment regimens follow similarly.

\subsection{Simulations on coverage probability} 
\label{apdx: simulation true PS}

Appendix Table \ref{tab: CI} shows the coverage probability of 95\% bootstrap confidence intervals in both the point treatment setting and the time-varying setting. Due to prolonged runtime, we only consider settings of $a = 1$ and $\bar{a} = (1,1)$.  Each setting uses bootstrap sampling with 2000 replicates for 1000 simulated datasets with $N=1500, 5000$. . All coverage probabilities are approximately 95\%, suggesting that bootstrap is a valid tool for statistical inference.

\begin{table}[H]
\renewcommand{\arraystretch}{1.2}
\begin{center}
\begin{tabular}{|c c c c c|}
 \hline
\multirow{2}*{} & \multirow{2}*{} & \multirow{2}*{Truth} & \makecell{Coverage probability \\ (estimated PS)} & \makecell{Coverage probability \\ (known PS)} \\
\hline
\multirow{2}*{$a = 1$}          & $N$ = 1500 & 0.915  & 95.4\% & 94.2\%  \\ 
                                & $N$ = 5000 &        & 95.5\%  & 95.1\% \\ 
\hline
\multirow{2}*{$\bar{a}=(1, 1)$} & $N$ = 1500 & 0.751 & 94.9\% &  95.0\% \\ 
                                & $N$ = 5000 &       & 94.2\%  &  94.9\%\\ 

\hline
\end{tabular}
\caption{Coverage probability of 95\% bootstrap confidence intervals in both the point treatment setting and the time-varying setting. Truth: true survival-incorporated median. PS: propensity score.}
\label{tab: CI}
\end{center}
\end{table}

\section{Additional results in LLFS application}

\subsection{Additional details of the LLFS application}
\label{apdx: applicaiton detials}
We calculated the adapted Framingham Risk Score \citeplatex{d2008general} based on the following formula:
\begin{align*}
        & Risk\ factor =  \ 3.06117 \ln(Age) + 1.12370 \ln(Total\ cholesterol)   \\ 
        & \qquad \qquad \qquad  - 0.93263 \ln (HDL\ cholesterol) + \mathbbm{1}_{Cigarette\ smoker} + \mathbbm{1}_{Diabetes} - 23.9802 \\
        & Adapted\ Framingham\ Risk\ Score =  100 \cdot (1- 0.88936^{\exp(risk\ factor)}).
\end{align*}
Compared to the original Framingham Risk Score, we excluded the term
\[
\ln(Systolic\ blood\ pressure) \times \mathbbm{1}_{On\ blood\ pressure\ medication}
\]
due to the lack of information on blood pressure medication usage in the LLFS.

We used the R function “weighted\_quantile” from the R package “MetricsWeighted” to estimate the survival-incorporated median. We used the R function ``boot" from the R package ``boot" to construct bootstrap CIs for the survival-incorporated median. 

\subsection{Baseline characteristics after IPTW}

\begin{table}[H]
\centering
\centerline{\includegraphics[width=3.2 in]{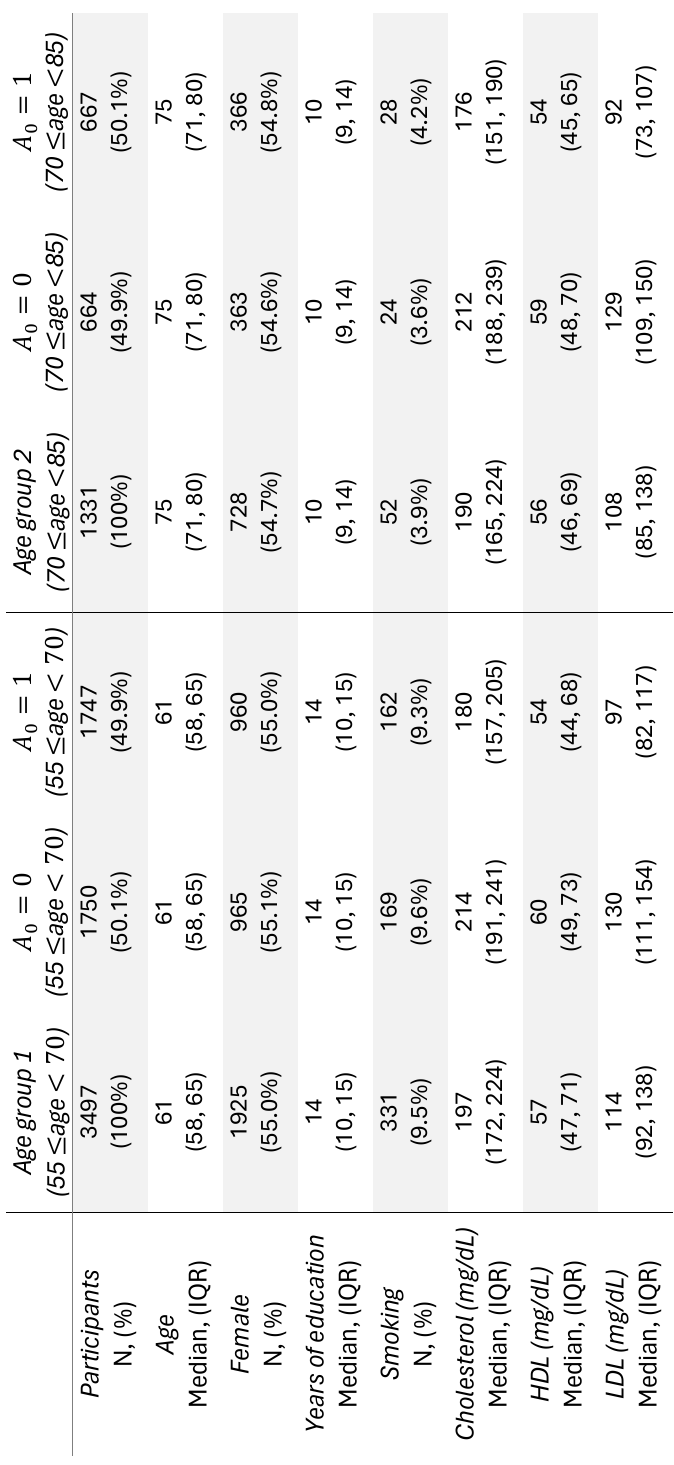}}

\caption{IPTW-weighted baseline characteristics of the Long Life Family Study participants; IPTW, Inverse Probability of Treatment Weighting (The IPTW is estimated based on a model of age and sex); N, number; IQR, interquartile range; HDL, high-density lipoprotein; LDL, low-density lipoprotein; mg/dL, milligrams per deciliter; $A_0 = 0$, participants off statins at baseline; $A_0 = 1$, participants on statins at baseline.}
\label{tab: weighted LLFS baseline}
\end{table}

\subsection{Results using the median in the survivors}
\label{apdx: applicaiton med survivors}

Table \ref{tab: LLFS med in sur res application} shows the estimated median cognitive change in the survivors in the LLFS application.
\begin{table}[H]
\renewcommand{\arraystretch}{1.25}
\begin{center}
\begin{tabular}{|p{1.6cm} c c|}
\hline
 & Treatment group & {Median in the survivors (95\% CI)}\\
 \hline
\multirow{3}{*}{age 55 - 69}  & $a=0$            &-3 [-4, -3]\\ 
                              & $a=1$            & -4 [-4, -2] \\ 
                              & $a=1$ - $a=0$    & -1 [-1, 2]\\
\hline
\multirow{3}{*}{age 70 - 84}  & $a=0$            & -5 [-6, -4]  \\ 
                              & $a=1$            & -5 [-7, -4] \\ 
                              &  $a=1$ - $a=0$   & 0  [-2, 3]\\

\hline
\end{tabular}
\caption{The median cognitive change of the DSST scores between 8 years and baseline in the survivors with bootstrap 95\% confidence intervals (CIs). Results are estimated with Inverse Probability of Treatment Weighting (IPTW) and Inverse Probability of Censoring Weighting (IPCW), restricting to only survivors. $a = 0$: participants off statins at baseline, had they remained off statins throughout. $a = 1$: participants on statins at baseline, had they remained on statins throughout. }
\label{tab: LLFS med in sur res application}
\end{center}
\end{table}

\bibliographystylelatex{plainnat}
\bibliographylatex{Bibliography.bib}

\end{document}